\documentclass[longauth, twocolumn]{aa}
\usepackage{graphicx}
\usepackage{enumitem}
\usepackage{txfonts}
\usepackage{xcolor}
\usepackage{xspace}
\usepackage{cancel}
\usepackage{ulem}
\usepackage[switch]{lineno}
\usepackage{orcidlink}
\usepackage{natbib}
\bibpunct{(}{)}{;}{a}{}{,} 

\newcommand{\bd}[1]{{\textbf{#1}}}
\newcommand{\dg}{$^{\circ}$\xspace}

\newcommand{\hess}{H.E.S.S.\xspace}
\newcommand{\sgra}{Sgr A$^\star$\xspace}

\newcommand{\hessGCsource}{HESS~J1745$-$290\xspace}

\newcommand{\ecut}{$E_{\rm{cut}}$\xspace}
\newcommand{\dfu}{TeV$^{-1}$\,cm$^{-2}$\,s$^{-1}$\xspace}

\begin{document} 

   \title{Evidence for a spectral steepening of the gamma-ray emission from the Galactic Center ridge}

    \titlerunning{The Galactic Center ridge $\gamma$-ray emission seen with H.E.S.S.}

    \author{\small
A.~Acharyya\inst{\ref{USD}}  \orcidlink{0000-0002-2028-9230} 
\and F.~Aharonian\inst{\ref{YSU},\ref{MPIK},\ref{DIAS}}  \orcidlink{0000-0003-1157-3915} 
\and H.~Ashkar\inst{\ref{LLR}}  \orcidlink{0000-0002-2153-1818} 
\and M.~Backes\inst{\ref{UNAM},\ref{NWU}}  \orcidlink{0000-0002-9326-6400} 
\and R.~Batzofin\inst{\ref{UP}}  \orcidlink{0000-0002-5797-3386} 
\and Y.~Becherini\inst{\ref{APC}}  \orcidlink{0000-0002-2115-2930} 
\and D.~Berge\inst{\ref{DESY},\ref{HUB}}  \orcidlink{0000-0002-2918-1824}
\and K.~Bernl\"ohr\inst{\ref{MPIK}}  \orcidlink{0000-0001-8065-3252} 
\and M.~B\"ottcher\inst{\ref{NWU}}  \orcidlink{0000-0002-8434-5692} 
\and C.~Boisson\inst{\ref{LUX}}  \orcidlink{0000-0001-5893-1797} 
\and J.~Bolmont\inst{\ref{LPNHE}}  \orcidlink{0000-0003-4739-8389} 
\and F.~Brun\inst{\ref{IRFU}}  \orcidlink{0000-0003-0770-9007} 
\and C.~Burger-Scheidlin\inst{\ref{DIAS}}  \orcidlink{0000-0002-7239-2248} 
\and T.~Bylund\inst{\ref{LUX}}  \orcidlink{0000-0003-2946-1313} 
\and S.~Casanova\inst{\ref{IFJPAN}}  \orcidlink{0000-0002-6144-9122} 
\and D.~Cecchin~Momesso\inst{\ref{ECAP}}  \orcidlink{0000-0001-6709-509X} 
\and M.~Cerruti\inst{\ref{APC}}  \orcidlink{0000-0001-7891-699X} 
\and M.~Chakraborty\inst{\ref{IRFU}}  
\and A.~Chen\inst{\ref{Wits}}  \orcidlink{0000-0001-6425-5692}
\and M.~Chernyakova\inst{\ref{DCU},\ref{DIAS}}  \orcidlink{0000-0002-9735-3608} 
\and J. O.~Chibueze\inst{\ref{NWU},\ref{UNAM}}  \orcidlink{0000-0002-9875-7436} 
\and O.~Chibueze\inst{\ref{NWU}}  \orcidlink{0000-0001-8601-2675} 
\and T.~Collins\inst{\ref{UP}}  \orcidlink{0000-0001-5020-5387}
\and B.~Cornejo\inst{\ref{IRFU}}  \orcidlink{0009-0003-0039-0483} 
\and G.~Cotter\inst{\ref{UOX}}  \orcidlink{0000-0002-9975-1829} 
\and G.~Cozzolongo\inst{\ref{ECAP}}  
\and J.~de~Assis~Scarpin\inst{\ref{LLR}}  \orcidlink{0009-0004-4411-236X}
\and M.~de~Naurois\inst{\ref{LLR}}  \orcidlink{0000-0002-7245-201X} 
\and E.~de~O\~na~Wilhelmi\inst{\ref{DESY}}  \orcidlink{0000-0002-5401-0744} 
\and J.~Devin\inst{\ref{LUPM}}\corrauth{devin@lupm.in2p3.fr} \orcidlink{0000-0003-1018-7246} 
\and A.~Djannati-Ata\"i\inst{\ref{APC}}  \orcidlink{0000-0002-4924-1708} 
\and A.~Dmytriiev\inst{\ref{Wits}}  \orcidlink{0000-0003-0102-5579} 
\and K.~Egberts\inst{\ref{UP}}  \orcidlink{0009-0000-5511-7060} 
\and K.~Egg\inst{\ref{ECAP}}  \orcidlink{0009-0002-4238-034X} 
\and C.~Esca\~{n}uela~Nieves\inst{\ref{MPIK}}  \orcidlink{0000-0002-7297-8126} 
\and P.~Fauverge\inst{\ref{LP2I}}  \orcidlink{0009-0006-1613-6633} 
\and S.~Fegan\inst{\ref{LLR}}  \orcidlink{0000-0002-9978-2510} 
\and K.~Feijen\inst{\ref{APC}}  \orcidlink{0000-0003-1476-3714} 
\and M.~D.~Filipovic\inst{\ref{Sydney}}  \orcidlink{0000-0002-4990-9288} 
\and G.~Fontaine\inst{\ref{LLR}}  \orcidlink{0000-0002-6443-5025} 
\and S.~Funk\inst{\ref{ECAP}}  \orcidlink{0000-0002-2012-0080} 
\and S.~Gabici\inst{\ref{APC}}  
\and Y.A.~Gallant\inst{\ref{LUPM}}  
\and M.~Genaro\inst{\ref{ECAP}}  \orcidlink{0000-0003-3461-1929} 
\and J.F.~Glicenstein\inst{\ref{IRFU}}  \orcidlink{0000-0003-2581-1742} 
\and J.~Glombitza\inst{\ref{ECAP}}  \orcidlink{0000-0001-9683-4568} 
\and P.~Goswami\inst{\ref{LSW}}  \orcidlink{0000-0001-5430-4374} 
\and M.-H.~Grondin\inst{\ref{LP2I}}  \orcidlink{0000-0002-8383-251X} 
\and L.~Heckmann\inst{\ref{APC}}  \orcidlink{0000-0002-6653-8407} 
\and B.~He{\ss}\inst{\ref{IAAT}}  \orcidlink{0009-0004-9999-171X} 
\and W.~Hofmann\inst{\ref{MPIK}}  \orcidlink{0000-0001-8295-0648} 
\and T.~L.~Holch\inst{\ref{DESY}}  \orcidlink{0000-0001-5161-1168}
\and M.~Holler\inst{\ref{Innsbruck}}  \orcidlink{0000-0002-0107-8657} 
\and M.~Jamrozy\inst{\ref{OAUJ}}  \orcidlink{0000-0002-0870-7778} 
\and F.~Jankowsky\inst{\ref{LSW}}  
\and A.~Jardin-Blicq\inst{\ref{LP2I}}  \orcidlink{0000-0002-6738-9351} 
\and I.~Jaroschewski\inst{\ref{IRFU}}  \orcidlink{0000-0001-5180-2845} 
\and D.~Jimeno\inst{\ref{DESY}}  \orcidlink{0009-0001-2499-9467} 
\and I.~Jung-Richardt\inst{\ref{ECAP}}  
\and K.~Katarzy{\'n}ski\inst{\ref{NCUT}}  \orcidlink{0000-0002-8806-4863}
\and D.~Kerszberg\inst{\ref{LPNHE}}  \orcidlink{0000-0002-5289-1509} 
\and B. Kh\'elifi\inst{\ref{APC}}  \orcidlink{0000-0001-6876-5577} 
\and N.~Komin\inst{\ref{LUPM},\ref{Wits}}  \orcidlink{0000-0003-3280-0582} 
\and K.~Kosack\inst{\ref{IRFU}}  \orcidlink{0000-0001-8424-3621} 
\and D.~Kostunin\inst{\ref{DESY}}  \orcidlink{0000-0002-0487-0076} 
\and R.G.~Lang\inst{\ref{ECAP}}  \orcidlink{0000-0003-0492-5628} 
\and S.~Lazarevi\'c\inst{\ref{Sydney}}  \orcidlink{0000-0001-6109-8548} 
\and A.~Lemi\`ere\inst{\ref{APC}}\corrauth{alemiere@apc.in2p3.fr} \orcidlink{0000-0002-6682-7188} 
\and M.~Lemoine-Goumard\inst{\ref{LP2I}}  \orcidlink{0000-0002-4462-3686} 
\and J.-P.~Lenain\inst{\ref{LPNHE}}  \orcidlink{0000-0001-7284-9220} 
\and P.~Liniewicz\inst{\ref{OAUJ}}  \orcidlink{0009-0008-3575-3965} 
\and A.~Luashvili\inst{\ref{NWU}}  \orcidlink{0000-0003-4384-1638} 
\and J.~Mackey\inst{\ref{DIAS}}  \orcidlink{0000-0002-5449-6131} 
\and D.~Maheso\inst{\ref{NWU}}  \orcidlink{0000-0003-0085-2820} 
\and D.~Malyshev\inst{\ref{IAAT}}  \orcidlink{0000-0001-9689-2194} 
\and D.~Malyshev\inst{\ref{ECAP}}  \orcidlink{0000-0002-9102-4854} 
\and V.~Marandon\inst{\ref{IRFU}}  \orcidlink{0000-0001-9077-4058} 
\and M.~G.~F.~Mayer\inst{\ref{ECAP}}  \orcidlink{0000-0002-9771-9841} 
\and A.~Mehta\inst{\ref{DESY}}  
\and A.M.W.~Mitchell\inst{\ref{ECAP}}  \orcidlink{0000-0003-3631-5648} 
\and R.~Moderski\inst{\ref{NCAC}}  \orcidlink{0000-0002-8663-3882} 
\and L.~Mohrmann\inst{\ref{MPIK}}  \orcidlink{0000-0002-9667-8654} 
\and H.~Ndiyavala\inst{\ref{UNAM},\ref{NWU}}  \orcidlink{0000-0001-9279-1775} 
\and J.~Niemiec\inst{\ref{IFJPAN}}  \orcidlink{0000-0001-6036-8569} 
\and P.~O'Brien\inst{\ref{Leicester}}  \orcidlink{0000-0002-5128-1899} 
\and L.~Olivera-Nieto\inst{\ref{GRAPPA}}  \orcidlink{0000-0002-9105-0518} 
\and M.O.~Moghadam\inst{\ref{UP}}  \orcidlink{0009-0003-2479-1863} 
\and S.~Panny\inst{\ref{Innsbruck}}  \orcidlink{0000-0001-5770-3805} 
\and M.~Panter\inst{\ref{MPIK}}  
\and R.D.~Parsons\inst{\ref{HUB}}  \orcidlink{0000-0003-3457-9308} 
\and P.~Pichard\inst{\ref{APC}}  \orcidlink{0009-0005-9803-0762} 
\and T.~Preis\inst{\ref{Innsbruck}}  \orcidlink{0009-0001-7110-6764} 
\and G.~P\"uhlhofer\inst{\ref{IAAT}}  \orcidlink{0000-0003-4632-4644} 
\and M.~Punch\inst{\ref{APC}}  \orcidlink{0000-0002-4710-2165} 
\and A.~Quirrenbach\inst{\ref{LSW}}  
\and A.~Reimer\inst{\ref{Innsbruck}}  \orcidlink{0000-0001-8604-7077} 
\and O.~Reimer\inst{\ref{Innsbruck}}  \orcidlink{0000-0001-6953-1385} 
\and Q.~Remy\inst{\ref{MPIK}}  \orcidlink{0000-0002-8815-6530} 
\and H.~X.~Ren\inst{\ref{MPIK}}  \orcidlink{0000-0003-0221-2560} 
\and B.~Reville\inst{\ref{MPIK}}  \orcidlink{0000-0002-3778-1432} 
\and F.~Rieger\inst{\ref{MPIK}}  \orcidlink{0000-0003-1334-2993} 
\and G.~Roellinghoff\inst{\ref{ECAP}}  \orcidlink{0000-0002-9824-9597} 
\and G.~Rowell\inst{\ref{Adelaide}}  \orcidlink{0000-0002-9516-1581} 
\and B.~Rudak\inst{\ref{NCAC}}  \orcidlink{0000-0003-0452-3805} 
\and K.~Sabri\inst{\ref{LUPM}}  
\and V.~Sahakian\inst{\ref{YPI}}  \orcidlink{0000-0003-1198-0043} 
\and A.~Santangelo\inst{\ref{IAAT}}  \orcidlink{0000-0003-4187-9560} 
\and M.~Sasaki\inst{\ref{ECAP}}  \orcidlink{0000-0001-5302-1866} 
\and F.~Sch\"ussler\inst{\ref{IRFU}}  \orcidlink{0000-0003-1500-6571} 
\and J.N.S.~Shapopi\inst{\ref{UNAM}}  \orcidlink{0000-0002-7130-9270} 
\and I.~Shebalkova\inst{\ref{DCU}}  
\and W.~Si~Said\inst{\ref{LLR}}  \orcidlink{0009-0007-6555-6893} 
\and H.~Sol\inst{\ref{LUX}}  
\and {\L.}~Stawarz\inst{\ref{OAUJ}}  \orcidlink{0000-0002-7263-7540} 
\and R.~Steenkamp\inst{\ref{UNAM}}  \orcidlink{0009-0009-4130-977X} 
\and S.~Steinmassl\inst{\ref{MPIK}}  \orcidlink{0000-0002-2865-8563} 
\and T.~Tanaka\inst{\ref{Konan}}  \orcidlink{0000-0002-4383-0368} 
\and A.M.~Taylor\inst{\ref{DESY}}  \orcidlink{0000-0001-9473-4758} 
\and G.~L.~Taylor\inst{\ref{LSW}}  \orcidlink{0009-0001-8062-036X} 
\and R.~Terrier\inst{\ref{APC}}\corrauth{rterrier@apc.in2p3.fr} \orcidlink{0000-0002-8219-4667} 
\and Y.~Tian\inst{\ref{DESY}}  \orcidlink{0009-0005-7165-3791} 
\and M.~Tsirou\inst{\ref{DESY}}  \orcidlink{0000-0003-3417-1425} 
\and N.~Tsuji\inst{\ref{ICRR}}  \orcidlink{0000-0001-7209-9204} 
\and T.~Unbehaun\inst{\ref{MPIK}}  \orcidlink{0000-0002-7378-4024}
\and C.~van~Eldik\inst{\ref{ECAP}} \orcidlink{0000-0001-9669-645X} 
\and C.~Venter\inst{\ref{NWU}}  \orcidlink{0000-0002-2666-4812} 
\and J.~Vink\inst{\ref{GRAPPA}}  \orcidlink{0000-0002-4708-4219} 
\and V.~Voitsekhovskyi\inst{\ref{GRAPPA}}  \orcidlink{0000-0002-3906-4840} 
\and T.~Wach\inst{\ref{MPIK}}  \orcidlink{0009-0008-4658-7405} 
\and S.J.~Wagner\inst{\ref{LSW}}  \orcidlink{0000-0002-7474-6062} 
\and A.~Wierzcholska\inst{\ref{IFJPAN},\ref{LSW}}  \orcidlink{0000-0003-4472-7204} 
\and M.~Zacharias\inst{\ref{LSW},\ref{NWU}}  \orcidlink{0000-0001-5801-3945} 
\and A.~Zech\inst{\ref{LUX}}  
\and W.~Zhong\inst{\ref{DESY}}  \orcidlink{0000-0003-3717-2861} 
}

\institute{
University of Southern Denmark (SDU), Campusvej 55, 5230 Odense, Denmark \label{USD}
\and Yerevan State University, 1 Alek Manukyan St, Yerevan 0025, Armenia \label{YSU}
\and Max-Planck-Institut für Kernphysik, P.O. Box 103980, D 69029 Heidelberg, Germany \label{MPIK}
\and Astronomy \& Astrophysics Section, School of Cosmic Physics, Dublin Institute for Advanced Studies, DIAS Dunsink Observatory, Dublin D15 XR2R, Ireland \label{DIAS}
\and Laboratoire Leprince-Ringuet, École Polytechnique, CNRS, Institut Polytechnique de Paris, F-91128 Palaiseau, France \label{LLR}
\and University of Namibia, Department of Physics, Private Bag 13301, Windhoek 10005, Namibia \label{UNAM}
\and Centre for Space Research, North-West University, Potchefstroom 2520, South Africa \label{NWU}
\and Institut für Physik und Astronomie, Universität Potsdam, Karl-Liebknecht-Strasse 24/25, D 14476 Potsdam, Germany \label{UP}
\and Université Paris Cité, CNRS, Astroparticule et Cosmologie, F-75013 Paris, France \label{APC}
\and Deutsches Elektronen-Synchrotron DESY, Platanenallee 6, 15738 Zeuthen, Germany \label{DESY}
\and Institut für Physik, Humboldt-Universität zu Berlin, Newtonstr. 15, D 12489 Berlin, Germany \label{HUB}
\and LUX, Observatoire de Paris, Université PSL, CNRS, Sorbonne Université, 5 Pl. Jules Janssen, 92190 Meudon, France \label{LUX}
\and Sorbonne Université, CNRS/IN2P3, Laboratoire de Physique Nucléaire, et de Hautes Energies, LPNHE, 4 place Jussieu, 75005 Paris, France \label{LPNHE}
\and IRFU, CEA, Université Paris-Saclay, F-91191 Gif-sur-Yvette, France \label{IRFU}
\and Instytut Fizyki Jac{a}drowej PAN, ul. Radzikowskiego 152, ul. Radzikowskiego 152, 31-342 Kraków, Poland \label{IFJPAN}
\and Friedrich-Alexander-Universität Erlangen-Nürnberg, Erlangen Centre for Astroparticle Physics,  Nikolaus-Fiebiger-Str. 2, 91058 Erlangen, Germany \label{ECAP}
\and School of Physics, University of the Witwatersrand, 1 Jan Smuts Avenue, Braamfontein, Johannesburg, 2050, South Africa \label{Wits}
\and School of Physical Sciences and Centre for Astrophysics \& Relativity, Dublin City University, Glasnevin, Dublin D09 W6Y4, Ireland \label{DCU}
\and University of Oxford, Department of Physics, Denys Wilkinson Building, Keble Road, Oxford OX1 3RH, UK, United Kingdom \label{UOX}
\and Laboratoire Univers et Particules de Montpellier, Université Montpellier, CNRS/IN2P3, CC 72, Place Eugène Bataillon, F-34095 Montpellier Cedex 5, France \label{LUPM}
\and Université Bordeaux, CNRS, LP2I Bordeaux, UMR 5797, F-33170 Gradignan, France \label{LP2I}
\and School of Science, Western Sydney University, Locked Bag 1797, Penrith South DC, NSW 2751, Australia \label{Sydney}
\and Landessternwarte, Universität Heidelberg, Königstuhl, D 69117 Heidelberg, Germany \label{LSW}
\and Institut für Astronomie und Astrophysik, Universität Tübingen, Sand 1, D 72076 Tübingen, Germany \label{IAAT}
\and Universität Innsbruck, Institut für Astro- und Teilchenphysik, Technikerstraße 25, 6020 Innsbruck, Austria \label{Innsbruck}
\and Obserwatorium Astronomiczne, Uniwersytet Jagielloński, ul. Orla 171, 30-244 Kraków, Poland \label{OAUJ}
\and Institute of Astronomy, Faculty of Physics, Astronomy and Informatics, Nicolaus Copernicus University, Grudziadzka 5, 87-100 Torun, Poland \label{NCUT}
\and Nicolaus Copernicus Astronomical Center, Polish Academy of Sciences, ul. Bartycka 18, 00-716 Warsaw, Poland \label{NCAC}
\and University of Leicester, School of Physics and Astronomy, University Road, Leicester, LE1 7RH, United Kingdom \label{Leicester}
\and GRAPPA, Anton Pannekoek Institute for Astronomy, University of Amsterdam, Science Park 904, 1098 XH Amsterdam, The Netherlands \label{GRAPPA}
\and School of Physical Sciences, University of Adelaide, Adelaide 5005, Australia \label{Adelaide}
\and Yerevan Physics Institute, 2 Alikhanian Brothers St., 0036 Yerevan, Armenia \label{YPI}
\and Department of Physics, Konan University, 8-9-1 Okamoto, Higashinada, Kobe, Hyogo 658-8501, Japan \label{Konan}
\and Institute for Cosmic Ray Research, University of Tokyo, 5-1-5, Kashiwa-no-ha, Kashiwa, Chiba 277-8582, Japan \label{ICRR}
}

   \date{Received 25 March 2026 / Accepted 11 August 2026}
 
  \abstract
   {Very-high-energy (VHE) $\gamma$-ray emission detected from the central molecular zone (CMZ) hints at diffusive propagation of cosmic rays (CRs) injected continuously near the Galactic center (GC).}
   {Using H.E.S.S. VHE $\gamma$-ray observations, we aim to construct a multi-component description of the region in order to derive the spatial and spectral distributions of CRs and discuss their potential origin.}
   {We rely on a spectro-morphological analysis, in which spectral and spatial parameters of different components are fit simultaneously, and on a three-dimensional description of the CMZ and CR distributions.}
   {The GC ridge emission is well reproduced with the assumed gas distribution model weighted by a $1/r^{\alpha}$ CR density profile with \bd{$\alpha = 1.10 \pm 0.05_{\rm{stat}} \pm 0.10_{\rm{syst}}$.} We detect no significant spectral variations across the CMZ, which supports a continuous injection from the GC. We report a significant (> 3 $\sigma$) curvature in the GC ridge $\gamma$-ray spectrum, implying a parent proton break (or cutoff energy) of $E_b \sim 15-50$~TeV ($E_{\rm{cut}} \sim 60-160$~TeV). We also derive the best-fit spectral parameters of the other sources in the region and report a significant cutoff at $E \sim 5$ TeV in the spectrum of the pulsar wind nebula G0.9+0.1. Finally we constrain the position of the injection site to be close to the GC (ruling out the Arches and Quintuplet clusters as major contributors to the emission). We discuss possible origins for the GC ridge emission, potential mechanisms for causing the observed curvature, a possible contribution of the gas content to the emission of \hessGCsource and derive qualitative constraints on the line-of-sight positions of the clouds shaping the CMZ.} 
   {}
   

   \maketitle

\nolinenumbers

\section{Introduction}

The Galactic center (GC) region harbors numerous potential sources of accelerated particles, such as supernova remnants (SNRs), pulsar wind nebulae (PWNe), young massive star clusters (Arches, Quintuplet and the Central clusters), and is immersed in a highly turbulent medium. \sgra, the supermassive black hole at the GC, has been actively monitored across the radio, X-ray, and infrared wavelengths for several decades, showing significant variability in all these bands \citep{Genzel2010}. With a mass of about $4 \times 10^6$\,M$_{\odot}$ \citep{EHT2022}, it presently shows a remarkably low accretion rate, typical of quiescent galactic nuclei \citep{Yuan2014}. However, evidence such as the detection of the Fermi bubbles \citep{Su2010} and X-ray light echoes \citep{Ponti:2010} suggests that \sgra experienced significantly higher activity in the past, and may have been an important site for the acceleration of very high-energy cosmic rays (CRs).

Different potential particle accelerators are embedded in the so-called central molecular zone (CMZ), a region resulting from inflow driven by the Galactic bar and containing $\sim$ 3$-$10$\%$ of the molecular gas in our galaxy while occupying only 1$\%$ of its volume \citep{Ferriere:2007}. Close to \sgra, the molecular gas is structured in the circumnuclear ring (CNR), a ring-like structure with an inner/outer radius of $\sim 1/7$\,pc \citep{Ferriere:2012, Oka:2011} and estimated densities of $n \sim 10^{5}-10^{7}$\,cm$^{-3}$. At larger scales the CMZ (with a mean density $n \sim 100$\,cm$^{-3}$) extends up to a Galactic radius $r \sim 200$\,pc, with an asymmetric distribution toward positive longitudes over the range of $l \approx$ ($-1.0$\dg, $+1.7$\dg) and a latitude range of $b$ $\approx$ ($-0.5$\dg, $+0.5$\dg). The CMZ region is mainly shaped by dense clouds with a total mass of about $(2-5) \times 10^{7}$\,M$_{\odot}$ \citep{battersby2025,Dahmen98}. Recent estimates, considering HI 21\,cm line emission and using several carbon monoxide isotopes, favor values at the lower end of this range, with $2.3 \times 10^{7}$\,M$_{\odot}$ \citep{Ren:2025}. Many of these clouds are identified and separated into structures such as the ones with local standard of rest velocities of 20\,km\,s$^{-1}$ and 50\,km\,s$^{-1}$, the Sgr~B complex (including Sgr~B2), the G1.3$^\circ$ cloud, the Dust Ridge and Sgr~C. Despite numerous studies, the three-dimensional geometry of the CMZ remains debated: although the gas in the CMZ is generally thought to be organized into a ring-like structure, its orientation, shape, individual cloud positions, and dynamics remain unclear \citep{Henshaw2023}.

Soon after a first hint of TeV emission towards the GC was found by the Whipple observatory \citep{Whipple_GC:2004}, the H.E.S.S. collaboration reported the detection of a very-high-energy (VHE) $\gamma$-ray point source, \hessGCsource, spatially coincident with the radio source \sgra within a positional uncertainty of 13" \citep{HESS_GC_detection:2004, HESS_GC_localization:2010}. \hessGCsource exhibits a hard spectrum, with a power-law index of $\Gamma \sim 2$, an exponential cut-off at around 10 TeV \citep{HESS_GC_source_spectrum:2009, HESS_GC_PeVatron:2016} and no variability detected on a daily or half-year timescale \citep{HESS_GC_source_spectrum:2009, VERITAS_GC:2021}. H.E.S.S. also reported TeV unresolved emission from the composite SNR G0.9+0.1 with a power-law spectrum \citep{HESS_G09:2005}, likely associated with the X-ray and radio PWN  \citep{Porquet:2003, G09_Xray_spec:2012}.

H.E.S.S. also detected diffuse VHE $\gamma$-ray emission along the CMZ (the GC ridge) which follows the dense gas distribution in that region, suggesting it is caused by hadronic interactions of multi-TeV CRs with gas in the GC massive clouds \citep{HESS_GC_ridge_discovery:2006}. In addition to the spatial correlation, the large extent of the $\gamma$-ray emission disfavors a leptonic origin due to severe radiative losses. The $\gamma$-ray brightness suggests an overdensity of CRs in the CMZ compared to the solar neighborhood, and thus could point towards the presence of at least one source of CRs in the region. Spatially resolved measurements have revealed that the CR density is not uniform, but is consistent with a $1/r$ profile ($r$ being the radial distance to the CR source), expected in a scenario in which a source in the vicinity of the GC is injecting CRs into the CMZ at a constant rate \citep{HESS_GC_PeVatron:2016, HESS_Diffuse_emission:2018, MAGIC_diffuse_GC:2020}. The supermassive black hole \sgra has been proposed as this potential steady CR source \citep{HESS_GC_PeVatron:2016}, but other sources are also suitable candidates, like a population of millisecond pulsars \citep{Guepin_Pevatron_ms_pulsars:2018}, SNRs \citep{Jouvin_SNR_CRs:2017, Jouvin_SNR_CRs:2020}, or stellar clusters \citep{Aharonian:2019}. The $\gamma$-ray spectrum of the GC diffuse emission can be fitted using a power-law model with an average spectral index of $\Gamma\sim 2.3$ \citep{HESS_GC_ridge_discovery:2006, HESS_GC_PeVatron:2016}. No evidence for a spectral curvature was observed neither in the central 0.45\dg nor on larger scales \citep{HESS_GC_PeVatron:2016, HESS_Diffuse_emission:2018}, with 250 hours of data extending over 10 years of H.E.S.S. observations. At large zenith angles, other Imaging Atmospheric Cherenkov Telescopes (IACTs), such as VERITAS and MAGIC, also observed the GC ridge and found a spectrum compatible with that from \hess \citep{Veritas_GC_ridge:2016, GC_MAGIC:2017, MAGIC_diffuse_GC:2020}, though a possible hint for a cutoff at $17.5_{-9.7}^{+7.4}$ TeV (at a $2\sigma$-confidence level) was indicated by MAGIC. At higher energies, HAWC measured a power-law spectral index of $\Gamma = 2.88 \pm 0.15$ from a point-like source near the GC, which is significantly softer than the one measured by H.E.S.S. and points therefore towards a spectral transition between the H.E.S.S. and HAWC energy ranges \citep{HAWC:2024_GC}. 

In this work, we revisit the GC region with H.E.S.S. using a deeper exposure and perform a spectro-morphological analysis in which the individual contributions from the different components are modeled and fitted simultaneously to the data.
This is particularly important for such an extended component as the GC ridge, for which the largest sources of systematic uncertainties arise from the modeling of the residual hadronic background and the Galactic large-scale emission. 
With a spectro-morphological $\gamma$-ray analysis applied at TeV energies, we are now able to fit in 3D($l$, $b$, $E$)\footnote{2D for space, 1D for energy.} multiple overlapping components and model the main sources of systematic uncertainties as nuisance parameters. In addition to this improvement over previous analyses, we use a three-dimensional description of the gas and CRs within the CMZ, and we test possible morphological and spectral deviations from the continuous injection scenario.  

We present H.E.S.S. observations and data reduction in Section~\ref{sec:data_reduction} followed by the details of the spectro-morphological analysis and the model used in Section~\ref{sec:spectro_ana}. Section~\ref{sec:HESS_sources} gives the best-fit spatial and spectral parameters of the point sources within this model. Section~\ref{sec:GC_ridge} is dedicated to the discussion of the GC ridge emission presenting the $\gamma$-ray spectrum and parent proton distribution. We also assess in Section~\ref{sec:steady} the validity of the steady-source scenario. Finally, Section~\ref{sec:discussion} provides discussions and interpretations of these results.

\section{H.E.S.S. observations and data reduction} \label{sec:data_reduction}

\hess is an array of five IACTs operating since 2003 and located in the Khomas Highland (Namibia) at 1800\,m above sea level. The initial four telescopes CT1$-$4 detect VHE $\gamma$-rays from $\sim$ 100\,GeV to several tens of TeV. In 2012, a fifth telescope (CT5) was added to the center of the array, lowering the energy threshold to $\sim 70$~GeV. The \hess point spread function (PSF), reaches its best value of 5' at multi-TeV energies. Due to its location in the Southern hemisphere, \hess is well suited to observe the GC. 

This study used data taken between March 2004 and June 2020, and reconstructed with the CT1$-$4. To reduce systematic uncertainties due to the absorption of Cherenkov light in the atmosphere, we limited our dataset to observations taken at zenith angles $<$ 40\dg, leading to a mean zenith angle of 18\dg. 
Furthermore, to minimize systematic uncertainties arising from the limited accuracy of the instrument response functions (IRFs), the analysis was restricted to events with an angular offset within 2\dg from the camera center and with reconstructed energies above the threshold at which the on-axis effective area reaches 10\% of its peak value.
This observation and photon selection resulted in a cumulated observation time of $\sim$ 365 hours in the entire CMZ region. Each calibrated run passing the quality criteria was analyzed with a configuration optimized for Galactic sources within the H.E.S.S. analysis package (HAP) framework described in \cite{Khelifi:2016}, including a Hillas-type shower reconstruction \citep{Hillas:1985} and a  multi-variate analysis technique \citep{Becherini:2012} for the $\gamma$-hadron discrimination. The results were cross-checked using independent methods for calibration, reconstruction \citep{Parsons:2014} and $\gamma$-hadron discrimination \citep{Ohm:2009}. The reconstructed and selected $\gamma$-like events as well as the IRFs were then exported into the GADF data format \citep{Deil:2017a}.

To be able to perform a complex spectro-morphological study, the data reduction and the following analysis were made with \textsc{gammapy} v1.2 \citep{gammapy_paper:2023, gammapy_1.2_zenodo}. We performed a binned analysis in a 6\dg $\times$ 4\dg region in Galactic coordinates, centered on the GC, using a spatial bin size of 0.02\dg and 0.2\dg for counts and IRF maps, respectively. This large region-of-interest allows better constraints on the Galactic large-scale emission, a dominant source of systematics for large extended sources as the GC ridge. During data reduction, we divided the reconstructed (true) energy into 30 (70) equally logarithmically-spaced energy bins from 200\,(80)\,GeV to 100\,(150)\,TeV. Only events with reconstructed photon energies $>400$\,GeV were kept for the analysis described here because effective area uncertainties are larger close to the energy threshold of the instrument.

IACT $\gamma$-ray measurements suffer from a residual background arising from showers initiated by CRs, being misclassified as $\gamma$ rays. The contribution of this residual background in the field must be accounted for in the spectro-morphological analysis and evaluated in each bin of the data cube in position and reconstructed energy $(l, b, E)$. Background models were derived from data in off-source regions (mainly using extragalactic observations) projected into a multi-dimensional table as function of observation conditions \citep{Mohrmann:2019, Abdalla_2021}. The estimation of the background for each run was then evaluated using a multi-variable interpolation of the model on the parameter space. Due to variations of observation conditions, the background level for a run can be affected by uncertainties in both normalization and spectral slope \citep[see e.g.][]{Mohrmann:2019}.
To mitigate the effects of these uncertainties on the astrophysical parameters extracted during the fit, the background template for each run was re-adjusted to observations, using nuisance parameters for the normalization in each reconstructed energy bin up to 8 TeV (followed by a power law), and excluding $\gamma$-ray bright regions following the field-of-view method described in \cite{Berge:2007}. Exclusion regions were defined around the Galactic plane ($\lvert b \rvert$ < 1.3\dg) and HESS~J1745$-$303 (using a circle of 0.73\dg centered on $l, b$ = 358.72\dg, $-0.87$\dg).

\section{Spectro-morphological analysis}\label{sec:spectro_ana}

In this section, the model components, the fit process and the method to derive the systematic uncertainties are described. The best-fit results are then given in Sections~\ref{sec:HESS_sources} and~\ref{sec:GC_ridge}.

    \subsection{Model components}\label{sec:models}

The model used in this work contains: four point sources previously reported by H.E.S.S., a template describing the $\gamma$-ray emission due to CR interactions within the CMZ, the Galactic large-scale emission and the residual hadronic background. Each of these components is described by a spatial model associated to a spectral shape. Throughout this paper, we use different spectral models, such as a power law (PL), a broken power law (BPL):
\begin{equation}\label{eq:BPL}
  \frac{dN}{dE} = N_0 \left\{
      \begin{aligned}
        \bigg (\frac{E}{E_{b}} \bigg) ^{-\Gamma_1} \hspace{0.5cm} \text{if} \hspace{0.5cm} E < E_b\\
        \bigg (\frac{E}{E_{b}} \bigg) ^{-\Gamma_2} \hspace{0.5cm} \text{if} \hspace{0.5cm} E > E_b \\
      \end{aligned}
    \right.
\end{equation}
a logarithmic parabola model (LogP):
\begin{equation}\label{eq:LP}
    \frac{dN}{dE} = N_0 \bigg ( \frac{E}{E_0} \bigg ) ^ {-\alpha - \beta \log(E/E_0)}
\end{equation}
and a power law with an exponential cutoff (ECPL):
\begin{equation}\label{eq:ECPL}
    \frac{dN}{dE} = N_0 \bigg ( \frac{E}{E_0} \bigg ) ^ {-\Gamma} \text{exp} \bigg ( -\frac{E}{E_{\rm{cut}}} \bigg )
\end{equation}
with $N_0$ the flux normalization (in units of TeV$^{-1}$\,cm$^{-2}$\,s$^{-1}$), $E_0$ the normalization energy (fixed to $1$\,TeV), $\Gamma_{(1,2)}$ (or $\alpha$, $\beta$) the spectral index (or spectral shape parameters) and $E_b$ (or \ecut) the break (or cutoff) energy (in TeV).

        \subsubsection{Known point sources}

We included in the model \hessGCsource, HESS~J1747$-$281 (the PWN in G0.9+0.1), HESS~J1741$-$302 and HESS~J1746$-$285 (also known as the `Arc source'). As a starting point for the modeling, we used their spatial and spectral parameters as reported in previous H.E.S.S. publications \citep{HESS_G09:2005, HESS_HESSJ1741:2018, HESS_Diffuse_emission:2018}. Therefore, the spectral shape of the sources is described by a PL except for \hessGCsource whose emission is described by an ECPL \citep{HESS_GC_PeVatron:2016}.

        \subsubsection{Galactic Center ridge template}\label{sec:template_GC_ridge}

To estimate the $\gamma$-ray distribution produced by the interaction of the CRs with the matter in the CMZ, we need to model the three-dimensional matter distribution in the region. Since the knowledge of the true gas kinematics in the GC is limited, we chose to base our gas distribution along the line of sight on the work of \cite{Sawada:2004}. Comparing the CO~2.6\,mm emission spectrum to the OH~18\,cm absorption towards the CMZ (at latitude $b$ = 0\dg), they extracted information on the gas distribution along the line of sight, and build a face-on mapping of the radial velocity as a function of longitude and line-of-sight distance which does not rely on any kinematic assumptions. We used this result and combined it with the CS ($l, b, v$) cube measured by \cite{Tsuboi:1999}, a good tracer of dense molecular structures, taking advantage of the good angular resolution of the CS survey, crucial for any morphological study. The velocity ranges were chosen to match those of \cite{Sawada:2004}, which are better-sampled in the central CMZ regions, where the most massive compact clouds are located, ensuring a more accurate mapping of cloud positions. We divided the CS intensity cube into 9 sub-cubes in the velocity ranges delimited by [$-200$, $-180$, $-140$, $-120$, $-70$, $-5$, $+40$, $+70$, $+120$, $+200$]\,km\,s$^{-1}$ and we assigned to each velocity band, for each longitude, the corresponding line-of-sight distance ($z$) predicted by \cite{Sawada:2004}. The obtained cube ($n_{gas}(x,y,z)$ thereafter) gives an approximate picture of the gas distribution in the GC region ($x$, $y$, $z$) = $400 \times 100 \times 1000$\,pc. This approach makes the assumption that CO and CS lines have similar radial velocity distributions in the CMZ, as demonstrated in \cite{RomanDuval2016}. It benefits from the high resolution of the CS data with a binning of less than 0.8\,pc in ($x$, $y$), but inherits the limited line-of-sight ($z$ axis) resolution of 30 pc from the initial CO map used by \cite{Sawada:2004}. This method has the advantage of not relying on gas dynamical models, although it inherits uncertainties from the assumptions made in \cite{Sawada:2004}. We note, in particular, that the matter distribution was originally derived at  $b = 0^\circ$ and is nonetheless applied here over all latitudes. However, as suggested by \cite{Yan_2017}, this assumption is overall not too far off, even though it remains approximate and does not fully capture the detailed latitude dependence.

We then modeled the spatial CR propagation within the CMZ assuming isotropic diffusion of protons from a source located at the GC. The CR density follows the  steady-state solution with a point-like continuous injection rate $\dot{N_p}$ and scales as the inverse of the distance to the injection point. We computed this distribution $n_{CR}(x,y,z)$ with an origin at the position of \sgra and with a similar binning to the one used for the gas distribution. We assumed that the diffusion regime starts at a minimal distance of 3\,pc (which corresponds to the spatial resolution used in this work of 0.02\dg), and extends up to a maximum distance of 500\,pc. Possible deviations from this stationary and isotropic solution are tested in Section~\ref{sec:steady}. 

Finally the $\gamma$-ray emission distribution was estimated by multiplying the CR distribution $n_{CR}(x,y,z)$ with that of the CMZ gas $n_{gas}(x,y,z)$. The integration of this cube along the line of sight allowed to obtain a 2D $\gamma$-ray spatial template that we used to model the morphology of the GC ridge emission.

        \subsubsection{Galactic large-scale emission}\label{sec:LSC}

H.E.S.S. previously demonstrated the existence of a diffuse large-scale VHE $\gamma$-ray emission along the Galactic plane \citep{Abramowski_2014,HGPS:2018}. The presence of this Galactic large-scale component complicates the extraction of the GC ridge emission, but with the 3D likelihood analysis, it can be treated as one of the $\gamma$-ray signal components with its specific spatial and spectral parametrizations. Following \cite{HGPS:2018}, we modeled this emission pragmatically as the product of a spectral component and a spatial component with uniform brightness along the Galactic longitude and a Gaussian latitudinal profile whose extent is left free. The Galactic large-scale emission was fitted simultaneously with the other components and several spectral shapes were tested: a PL, a LogP and an ECPL. We verified that using different spatial and spectral models does not affect our main conclusions (details are given in Appendix~\ref{appendix:syst_spatial_spec_models}) and we treated this component as a nuisance parameter.

    \subsection{Additional nuisance parameters and full model convolved with the IRFs}\label{sec:nuisance_param}

The spectral fit is particularly sensitive to systematic uncertainties in the effective area near the energy threshold, as well as in the energy distribution of the background model, both of which can bias the estimated spectral index and affect the measurement of a potential spectral curvature. We therefore introduced nuisance parameters in the lowest energy bin of the effective area, and in each bin of reconstructed energy of the background spectral model during the fit\footnote{For a comprehensive study on the treatment of systematic uncertainties using nuisance parameters and priors, see \cite{Katrin2025_prep}.}. Monte-Carlo simulations were performed to ensure an adequate estimation of the source spectrum with this method and showed that it can correct for background mismodeling up to 15\% in global normalization and 0.1 in spectral slope (more details are given in Appendix~\ref{appendix:syst_MC}).

The full model incorporates point sources, extended emissions (the GC ridge and the Galactic large-scale component), convolved with the IRFs, and the residual hadronic background model, along with the associated nuisance parameters described above.

   \subsection{Fit process and contribution of the different components}\label{sec:contribution_comp}

\begin{figure}[t]
    \centering
    \includegraphics[scale=0.35]{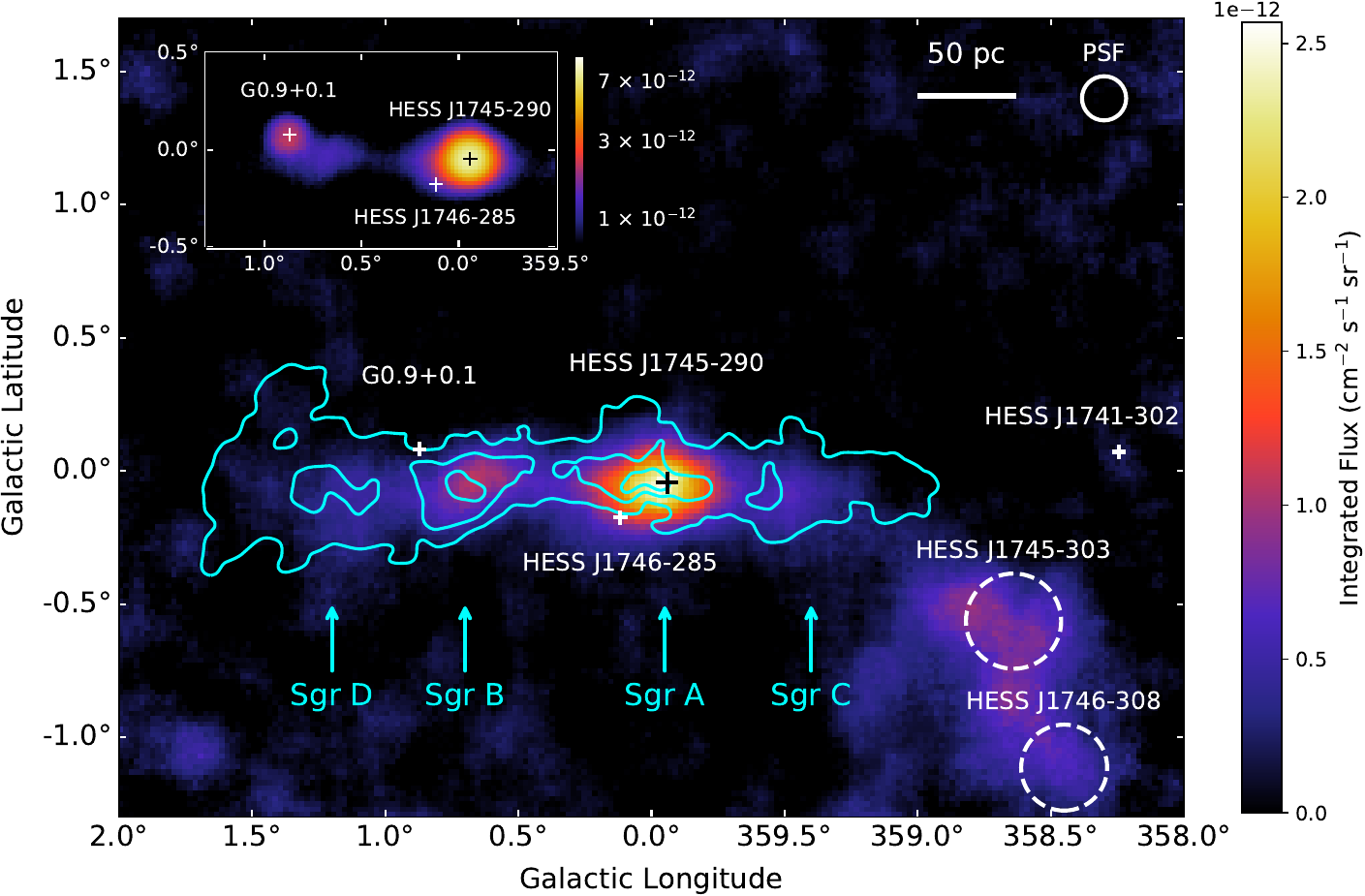}
    \caption{H.E.S.S. flux map ($E$ > 400 GeV) obtained after subtracting the contributions from \hessGCsource, G$0.9+0.1$ and the Galactic large-scale emission. The map is built with a correlation radius of 0.1\dg and assuming power-law models with $\Gamma = 2$. Cyan contours depict the CMZ, as seen with CS line emission \citep{Tsuboi:1999} and shaped by the cloud complexes Sgr~A, B, C and D. The position of the point sources in the model are represented by white crosses (black for \hessGCsource). For the analysis, a region encompassing HESS~J1745$-$303 and HESS~J1746$-$308 is masked (see the text). The inset plot shows the total flux towards the Galactic center (i.e. without subtracting sources) in logarithmic scale.}
    \label{fig:flux_map}
\end{figure}

We first iteratively fitted the position of the point sources, together with their spectral parameters, leaving $\sigma$, the Galactic large-scale emission latitudinal extent, free. We then fixed their spatial parameters as well as the large-scale emission latitudinal extent and tested different spectral shapes for the diffuse emissions (GC ridge and Galactic large-scale emissions) and derived the best-fit spectral parameters. We then refitted the spectral parameters of all the components together. 

Figure~\ref{fig:flux_map} depicts the H.E.S.S. flux map after subtracting the fitted contributions of the two brightest sources (\hessGCsource and G0.9+0.1 shown in the inset plot) and the Galactic large-scale emission. Bright $\gamma$-ray emission is visible within the CMZ (outlined by the CS gas contours), the so-called GC ridge, peaking towards the GC. For the analysis, we masked the complex region of HESS~J1745$-$303 and HESS~J1746$-$308, using a circle of 0.73\dg centered on $l, b$ = 358.72\dg, $-0.87$\dg. 

\begin{figure*}[t!]
    \centering
    \begin{minipage}{.5\textwidth}
        \centering
        \includegraphics[scale=0.6]{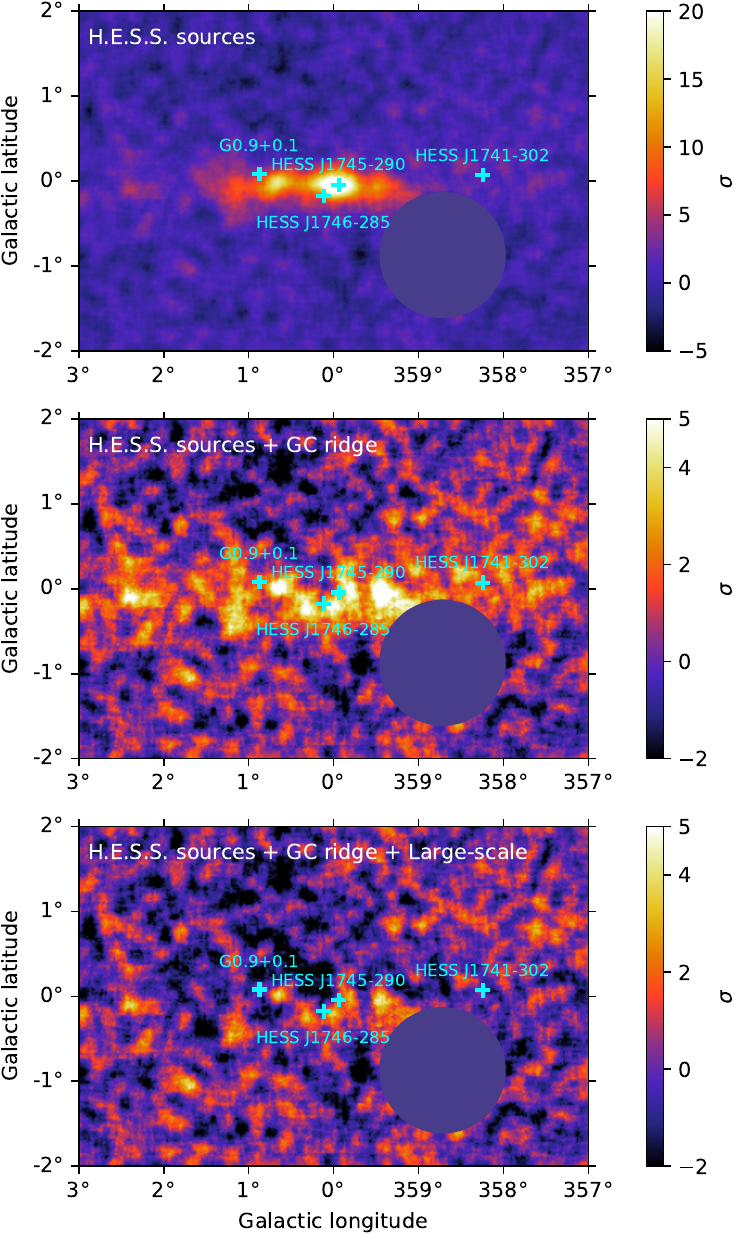}
    \end{minipage}
    \begin{minipage}{0.49\textwidth}
        \centering
        \includegraphics[scale=0.42]{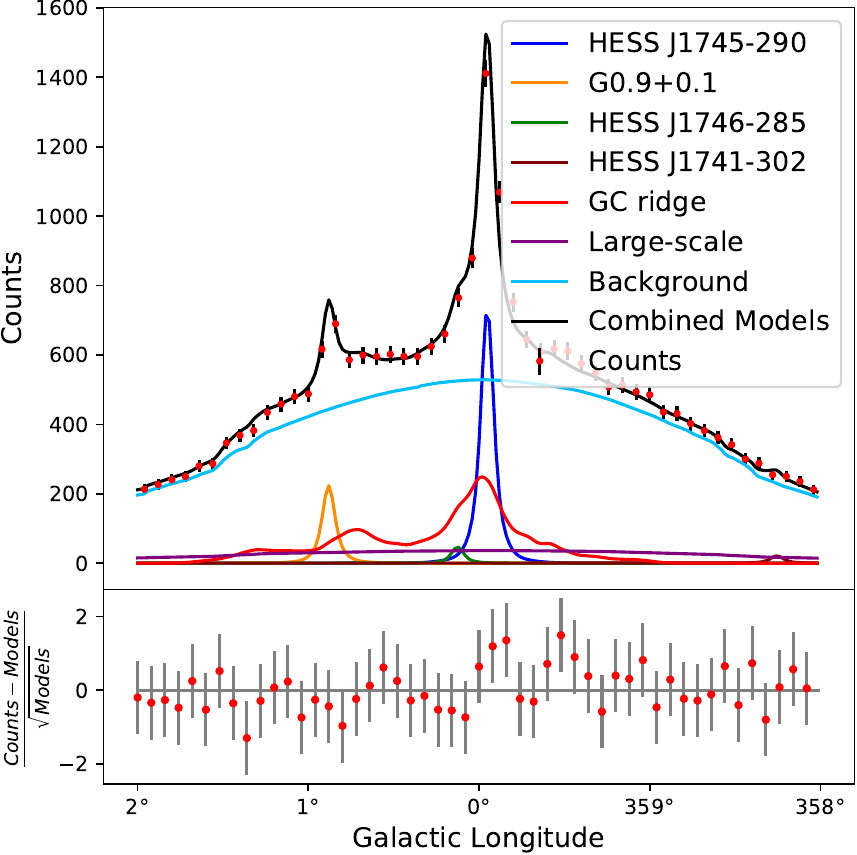}
        \includegraphics[scale=0.42]{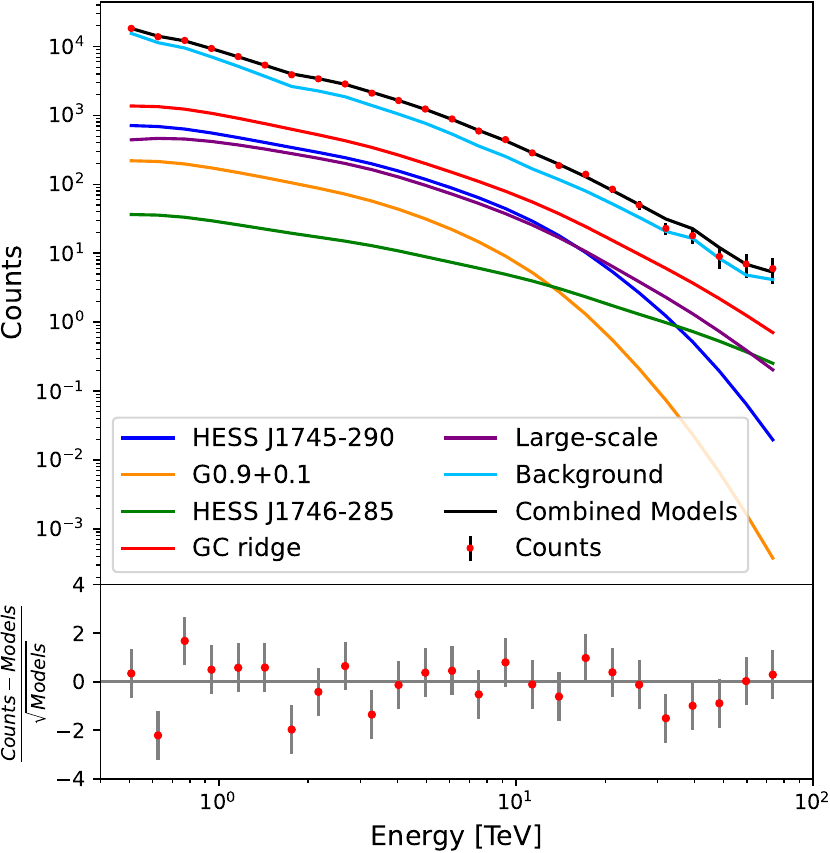}
    \end{minipage}
    \caption{(Left) Residual significance maps from 400 GeV to 100 TeV obtained after including the different components. The masked region is excluded from the analysis. (Right) Predicted longitude count profiles (top) of each component (colored lines) in the entire 6\dg $\times$ 4\dg region and corresponding spectral counts (bottom) in a 2\dg $\times$ 0.5\dg region around the CMZ (therefore excluding HESS J1741$-$302). The measured counts (red points) are reproduced by the full model described in this work (black line). We note that the exposure is not uniform along the longitude.}
    \label{fig:Fit_results}
\end{figure*}

Figure~\ref{fig:Fit_results} (left) shows the residual significance maps when including iteratively in the model the point sources, the GC ridge and the Galactic large-scale component. Adding the Galactic large-scale component improves the fit by $\Delta$TS = 466.0. We found a best-fit spatial extent of $\sigma_{\rm{lat}}$ = 0.32\dg $\pm$ 0.02\dg, which is compatible with the values reported in \cite{HGPS:2018} ranging typically between 0.2\dg and 0.3\dg over $\pm10$\dg in Galactic longitude, with an average error of 0.05\dg. This demonstrates that the Galactic large-scale component significantly contributes to the observed $\gamma$-ray emission, as also indicated by the residual positive excesses in the middle panel of Figure~\ref{fig:Fit_results} (left). When fitting the spectrum by a PL, we found a spectral index of $\Gamma = 2.31 \pm 0.05$, harder than what would be expected from local measurement but consistent with other experiments \citep{Gaggero:2015, Yang:2016}. Testing for different spectral shapes, the data indicate a clear preference for spectral curvature of the Galactic large-scale component: a LogP or an ECPL model is significantly preferred to the PL model, with a best-fit cutoff at several TeV. We note that the Galactic diffuse emission reported by LHAASO exhibits a spectral index of $\Gamma \sim $ 3 \citep{LHAASO_GIEM:2023}, which is steeper than the one expected from the local CR spectrum. A curvature or break in the Galactic large-scale emission might therefore be a common feature in the Galaxy. We also note that the large-scale component measured in this work is likely a combination of diffuse emission from Galactic CRs (hadronic and leptonic) and emission from unresolved sources\footnote{A detailed study of the Galactic CR sea diffuse emission with H.E.S.S. will be presented in a forthcoming publication.}. Because LHAASO and TibetAS$\gamma$ revealed Galactic diffuse emission which extends up to 100~TeV \citep{Amenomori:2021, LHAASO_GIEM:2023}, we used a LogP model throughout this paper. The fit of our final model gave no significant residual excess (Figure~\ref{fig:Fit_results}, bottom left). Figure~\ref{fig:Fit_results} (top right) shows the estimated longitude count profiles of all model components in the entire region (6\dg $\times$ 4\dg), the sum of which matches the data well. The spectral counts attributed to each component in the CMZ region (within a rectangular box of 2\dg $\times$ 0.5\dg) are given in Figure~\ref{fig:Fit_results} (bottom right), highlighting that the best-fit spectral shapes reproduce the observed counts well. 

    \subsection{Systematic uncertainties}\label{sec:Syst_uncertainties}

Systematic uncertainties in the IRFs are unavoidable, as the latter are usually derived for each run by interpolating over a predefined grid of observational conditions, which cannot fully capture the variability of the actual observing environment \citep{Bernlohr2008}, e.g. variations in telescope optical efficiency or atmosphere transparency which affect reconstructed $\gamma$-ray energies and effective area \citep{Hahn:2014}. While simulations and analysis aim to account for the variations in instrument parameters, atmosphere, and to properly model the residual hadronic background spectrum (e.g. through the introduction of nuisance parameters), some residual effects may remain. We therefore performed Monte-Carlo simulations to evaluate the impact of these uncertainties on the best-fit spectral parameters of the H.E.S.S. sources and of the GC ridge (Sections~\ref{sec:HESS_sources} and~\ref{sec:GC_ridge}). Using \textsc{gammapy}, we introduced several systematic effects on the IRFs and background model in the simulations of fake event cubes $(l, b, E)$, and refitted our best-fit model on these simulated data cubes to estimate residual systematic errors on the spectral parameters. The methodology is described in detail in Appendix~\ref{appendix:syst_MC} and the resulting parameter uncertainties are reported in Table~\ref{tab:best_fit_spec} and Table~\ref{tab:CMZ_spec}.

\section{Known point sources}\label{sec:HESS_sources}

\begin{table*}[th!]
    \caption{Best-fit spatial parameters of the point sources with associated statistical (1$\sigma$) errors.}
    \label{tab:best_fit_pos}
    \centering
    \begin{tabular}{l|ccccc}
    	\hline
        \hline
        Source  & $l$ & $b$  \\
		\hline
     HESS J1745$-$290 & 359.941\dg $\pm$ 0.001\dg &	$-0.043$\dg $\pm$ 0.001\dg  \\
     G0.9+0.1 & $0.872$\dg $\pm$ 0.002\dg & $0.082$\dg $\pm$ 0.002\dg \\
     HESS J1746$-$285 & $0.116$\dg $\pm$ 0.008\dg & $-0.163$\dg $\pm$ 0.008\dg \\
     HESS J1741$-$302 & 358.244\dg $\pm$ 0.011\dg & 0.072\dg $\pm$ 0.012\dg \\
     \hline
    \end{tabular}
    \vspace{0.05cm}
\end{table*}

\begin{table*}[th!]
    \caption{Best-fit spectral parameters of the point sources with associated statistical and systematic uncertainties.}
    \label{tab:best_fit_spec}
    \centering
    \begin{tabular}{l|ccc}
    	\hline
        \hline
        Source  & $N_0$ (\dfu) & $\Gamma$ & \ecut (TeV)   \\
		\hline
     HESS J1745$-$290 & $(1.89 \pm 0.05 \pm 0.39_{\rm{syst}}) \times 10^{-12}$ & $1.97 \pm 0.04 \pm 0.10_{\rm{syst}}$ & $9.38 \pm 1.05 \pm 1.30_{\rm{syst}}$ \\
     G0.9+0.1  &  $(7.55 \pm 0.72 \pm 1.45_{\rm{syst}} ) \times 10^{-13}$  & $1.87 \pm 0.15 \pm 0.09_{\rm{syst}} $ & $5.04_{-1.26}^{+2.48} \pm 0.50_{\rm{syst}}$ \\
     HESS J1746$-$285 & $(9.73 \pm 1.69 \pm 1.60_{\rm{syst}}) \times 10^{-14}$  & $1.99 \pm 0.12 \pm 0.15_{\rm{syst}}$ & $-$ \\
     HESS J1741$-$302 & $(10.08 \pm 2.34 \pm 1.97_{\rm{syst}}) \times 10^{-14}$ & $2.12 \pm 0.18 \pm 0.07_{\rm{syst}}$ & $-$ \\
     \hline
    \end{tabular}
    \vspace{0.05cm}
\end{table*}

The best-fit spatial and spectral parameters of the known point sources are reported in Tables~\ref{tab:best_fit_pos} and ~\ref{tab:best_fit_spec} respectively. We found compatible position with respect to published results within 1$\sigma$ statistical uncertainties and a pointing uncertainty of H.E.S.S. of roughly 20''$-$30'' \citep{Gillessen:2003}. We note that the best-fit position of HESS~J1746$-$285 was previously more shifted towards the North-East with $l$ = 0.14\dg $\pm$ 0.01$^{\circ}_{\rm{stat}}$ $\pm$ 0.01$^{\circ}_{\rm{syst}}$ and $b$ = $-0.11$\dg $\pm$ 0.02$^{\circ}_{\rm{stat}}$ $\pm$ 0.02$^{\circ}_{\rm{syst}}$ \citep{HESS_Diffuse_emission:2018} but the one derived in this work stays compatible with the position of the GC radio Arc and the X-ray PWN G0.13$-$0.11. Investigating the morphology of these four sources is out of the scope of this paper, as it would require a detailed analysis of the IRFs and PSF-related systematics.

The best-fit spectrum of \hessGCsource is compatible with the results given in \cite{HESS_GC_PeVatron:2016}. The emission of the PWN G0.9+0.1 was initially described by a PL \citep{HESS_G09:2005}, but with this significantly larger data set an ECPL model improves the fit by $\Delta$TS = 18.7 (a LogP model is similarly preferred with $\Delta$TS = 18.5), which corresponds to a significance of $\sim 4.3 \sigma$. We therefore detect a significant cutoff energy at $E_{\rm{cut}} = 5.04_{-1.26}^{+2.48} \pm 0.50_{\rm{syst}}$ TeV. The emissions of HESS~J1746$-$285 (the Arc source) and HESS~J1741$-$302 are described by a PL and are compatible with previous publications \citep{HESS_Diffuse_emission:2018, HESS_HESSJ1741:2018}. 

\begin{figure}[h!]
    \centering
    \includegraphics[scale=0.38]{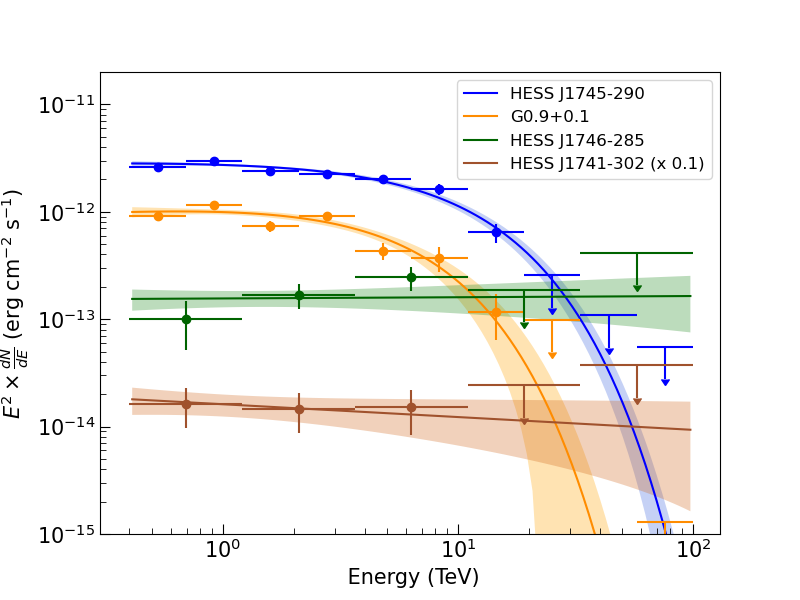}
    \caption{Best-fit spectra and SEDs with associated statistical uncertainties (1$\sigma$), and upper limits computed at a 2$\sigma$-confidence level. For visibility purpose, the emission of HESS~J1741$-$302 is scaled by a factor of 0.1.}
    \label{fig:spectra_HESS_sources}
\end{figure}

Using the best-fit model, we computed the spectral energy distributions (SEDs) of these sources, dividing the whole energy range into 4 logarithmically-spaced energy bins per decade. In each energy bin, the amplitude of the source of interest and the normalization of the background were allowed to vary during the fit. The statistical errors and upper limits were computed at a 1$\sigma$ and 2$\sigma$-confidence levels, respectively. For HESS~J1741$-$302 and HESS~J1746$-$285, we merged some energy bins due to low statistics. Figure~\ref{fig:spectra_HESS_sources} shows the corresponding SEDs and best-fit spectra with the statistical uncertainties. 

\section{The Galactic Center ridge}\label{sec:GC_ridge}

We used the spatial template described in Section~\ref{sec:template_GC_ridge} and tested different spectral models (PL, BPL, LogP, and ECPL) whose best-fit values are reported in Table~\ref{tab:CMZ_spec}. The PL spectral parameters are compatible with those obtained in \cite{HESS_GC_PeVatron:2016} and \cite{HESS_Diffuse_emission:2018}. However, owing to increased statistics and to the spectro-morphological analysis which allows multiple emission components and the main sources of systematic uncertainties to be modeled simultaneously, we now detect a significant curvature ($> 3\sigma$) in the GC ridge spectrum. The corresponding $\Delta$TS values (with respect to the PL model), given in Table~\ref{tab:CMZ_spec}, show that the BPL, the LogP and the ECPL are equally preferred in the 0.4$-$100 TeV range. For the BPL model we had to fix either $E_{\rm{b}}$ or $\Gamma_2$ to ensure convergence. We fixed $\Gamma_2 = 2.88$, the spectral index value reported by HAWC toward the GC \citep{HAWC:2024_GC} (fixing $\Gamma_2 = 3$ gives similar results: $\Gamma_1$ = 2.23 $\pm$ 0.04 and $E_{\rm{b}}$ = 6.34 $\pm$ 0.20\,TeV). We note that we obtained a higher significance value for the curvature ($> 5\sigma$) and a spectral transition at lower energy for the GC ridge when not including a curvature in the large-scale emission (clearly required by our data), as shown in Appendix~\ref{appendix:syst_spatial_spec_models}. Figure~\ref{fig:CMZ_gammaray_spectrum} (left) depicts the best-fit models and corresponding SEDs, showing that the curved spectra are similar up to $\sim 30$\,TeV, an energy at which the H.E.S.S. sensitivity decreases; the models therefore cannot be distinguished because of limited statistics. However, HAWC reported $\gamma$-ray emission extending up to 100\,TeV, and if indeed associated with the GC ridge emission, this would favor a curved model (a LogP or a BPL) rather than a cutoff. For illustration, Figure~\ref{fig:CMZ_gammaray_spectrum} (right) shows the GC ridge spectrum derived in this work with the ones obtained by MAGIC, VERITAS and HAWC. Although the spectra are consistent, we note that the comparison should be made with caution because extraction regions, analysis methods and modeling differ. 

\begin{table*}[t!]
     \caption{Best-fit spectral parameters for the GC ridge $\gamma$-ray emission with associated statistical (1$\sigma$) and systematic errors.}
    \label{tab:CMZ_spec}
    \centering
    \footnotesize
    \begin{tabular}{l|cccc|c}
    	\hline
        \hline
        Model& $N_0$ (\dfu) & $\Gamma$, $\Gamma_1$ or $\alpha$  & $\Gamma_2$ or $\beta$ & $E_{\rm{b}}$ or \ecut (TeV) &   $\Delta$TS \\
		\hline
        PL & $(3.89 \pm 0.15 \pm 0.92_{\rm{syst}}) \times 10^{-12}$ & $2.35 \pm 0.04 \pm 0.10_{\rm{syst}}$ & $-$ & $-$ & 0 \\
        BPL & $(3.86 \pm 0.10 \pm 0.44_{\rm{syst}}) \times 10^{-12}$ & $2.24 \pm 0.05 \pm 0.10_{\rm{syst}}$ & 2.88 (fixed) & $6.30 \pm 0.22 \pm 0.72_{\rm{syst}}$ & 11.5 \\
        LogP & $(4.00 \pm 0.10 \pm 0.76_{\rm{syst}}) \times 10^{-12}$ & $2.14 \pm 0.06 \pm 0.11_{\rm{syst}}$ & $0.12 \pm 0.03 \pm 0.05_{\rm{syst}}$ & $-$ & 14.1\\
        ECPL & $(4.11 \pm 0.18 \pm 0.93_{\rm{syst}}) \times 10^{-12}$ & $2.14 \pm 0.06 \pm 0.10_{\rm{syst}}$ & $-$ & $18.42_{-4.53}^{+9.13} \pm 3.20_{\rm{syst}}$ & 14.5 \\
     \hline
    \end{tabular}
    \vspace{0.05cm}
    \tablefoot{The $\Delta$TS value is given with respect to the PL model and the flux normalization $N_0$ is evaluated at $E_0$ = 1\,TeV for each model.}
\end{table*}

\begin{figure*}[t!]
	\centering
    \includegraphics[scale=0.36]{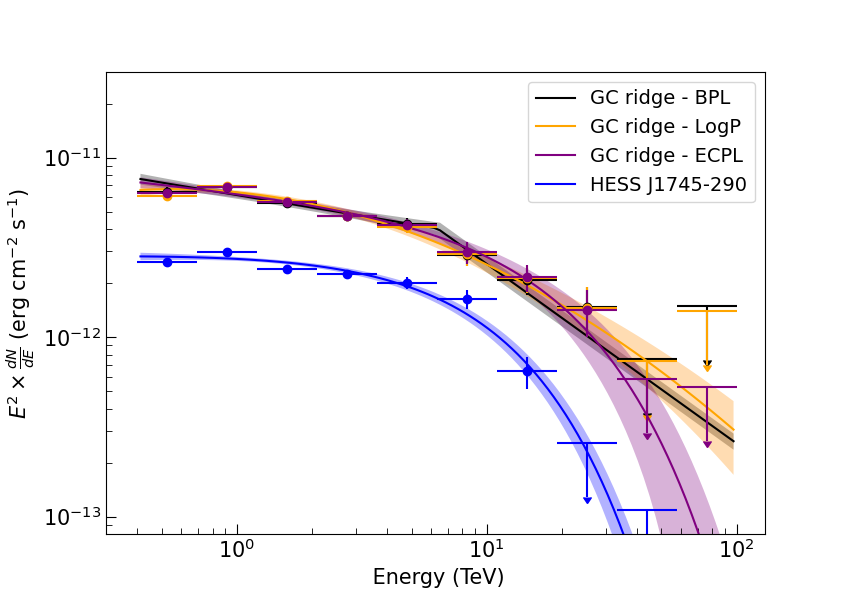}
    \includegraphics[scale=0.36]{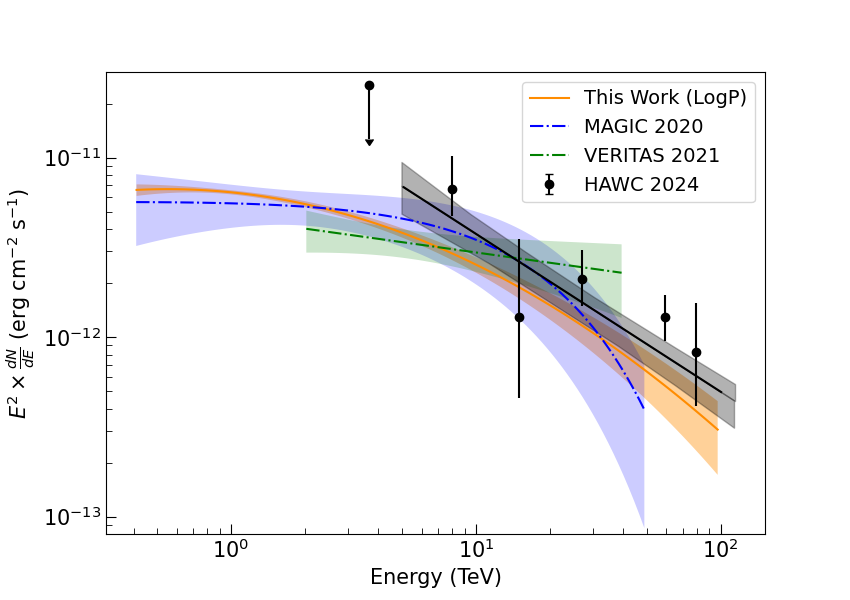}
    \caption{(Left) Best-fit spectrum and SEDs of the GC ridge (with that of \hessGCsource) with the associated statistical errors (1$\sigma$). The upper limits are computed at a 2$\sigma$-confidence level.  The SED points are derived assuming the corresponding best-fit spectral shape. (Right) Comparison of the logarithmic parabola model obtained in this work with the VERITAS, MAGIC and HAWC spectra extracted towards the GC region (with different extraction regions).}
    \label{fig:CMZ_gammaray_spectrum}
\end{figure*}

\begin{table*}[t!]
    \caption{Best-fit parent proton spectral parameters for the GC ridge with associated statistical (1$\sigma$) errors and taking $E_{\rm{p,0}}=10$\,TeV, $n_{cst}$ = 100\,cm$^{-3}$ and $d = 8.2$\,kpc.}
    \centering
    \footnotesize
    \begin{tabular}{l|cccc}
    	\hline
        \hline
        Model& $N_{p,0}$ (eV$^{-1}$) & $\Gamma$, $\Gamma_1$ or $\alpha$  & $\Gamma_2$ or $\beta$ & $E_{\rm{b}}$ or \ecut (TeV) \\
		\hline
        BPL & $(1.06 \pm 0.08) \times 10^{34}$ & $2.10 \pm 0.16$ & 2.9 (fixed) & $31.8 \pm 14.8$ \\
        LogP & $(1.17 \pm 0.07) \times 10^{34}$ & $2.15 \pm 0.10$ & $0.15 \pm 0.05$ & $-$ \\
        ECPL & $(1.17 \pm 0.09) \times 10^{34}$ & $2.04 \pm 0.13$ & $-$ & $111.4 \pm 50.8$ \\
     \hline
    \end{tabular}
    \vspace{0.05cm}
    \label{tab:CMZ_proton_spec}
    \tablefoot{The pion decay implementation follows \citet{Kafexhiu:2014} and contains a nuclear enhancement factor.}
\end{table*}

The best-fit parameters of the parent proton population were derived with the \texttt{naima} library \citep{Zabalza:2015} following \citet{Kafexhiu:2014} parametrization of pion decay emission. They are given in Table~\ref{tab:CMZ_proton_spec} assuming a constant mean target density of $n_{cst}$ = 100\,cm$^{-3}$ at a distance $d = 8.2$\,kpc. From the fitted normalization of the GC ridge template, we also derived a typical CR energy density profile above 10\,TeV of (details are given in Appendix~\ref{app:proton_distrib}):
$$w_{CR}(>10 \rm\ TeV) \sim 0.24 \pm 0.05\ eV\,cm^{-3} \left(\frac{r}{3\ pc}\right)^{-1}$$
Integrating it over a cylindrical region of radius 200\,pc and half-height 30\,pc,  yields a total proton energy of $W_p$ ($E_{p} >$ 10\,TeV) $\sim 3.5 \times 10^{48}$\,erg, which is less than the value reported in \cite{HESS_GC_PeVatron:2016} and \cite{HESS_Diffuse_emission:2018} mainly due to the inclusion of the Galactic large-scale component in the modeling and the detected curvature \footnote{We note however that these values cannot be directly compared because they were derived in different integration regions and with a different assumption for the nuclear enhancement factor: $\eta \sim 1.5$ and $\eta \sim 1$ in \cite{HESS_GC_PeVatron:2016} and \cite{HESS_Diffuse_emission:2018} respectively.}.

\hessGCsource is best-fitted with an ECPL rather than a LogP spectrum, but since we detect a curvature or a cutoff in the GC ridge spectrum, we tested the similarity of their spectra. Tying their ECPL spectral shapes (leaving free their amplitude), the fit degrades by $\Delta$TS = $-2.1$ with respect to the model with independent ECPL spectra, indicating that the spectrum of the GC ridge and \hessGCsource are not significantly different. We note that these spectra could be even more similar in terms of best-fit curvature or cutoff energy if the one of \hessGCsource is affected by $\gamma$-$\gamma$ absorption with intense radiation fields at the GC (this estimate is however beyond the scope of this paper).

\section{Testing the validity of the steady-source scenario}\label{sec:steady}

    \subsection{CR density profile}

\begin{figure}[t!]
	\centering
    \includegraphics[scale=0.45]{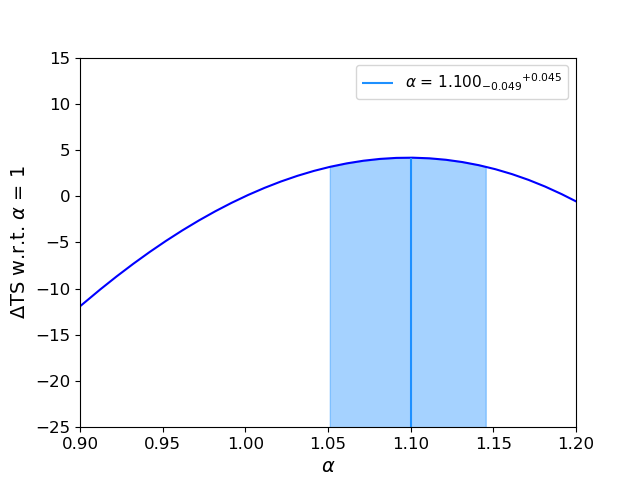}
 \caption{Likelihood profile (with respect to $\alpha = 1$) of the best-fit radial index obtained using a three-dimensional geometry of the CS gas.}
 \label{fig:morpho_CMZ}
\end{figure}

Throughout the analysis, we used a CR density profile $\propto 1/r^{\alpha}$ with $\alpha=1$ as indicated by previous studies. The 1D fit of the CR density profile on H.E.S.S. data gave $\alpha = 1.10 \pm 0.12$ \citep{HESS_GC_PeVatron:2016} while MAGIC reported a value of $\alpha = 1.2 \pm 0.3$, from a 2D fit \citep{MAGIC_diffuse_GC:2020}, i.e. the surface brightness $S(x, y) \approx A \int_{}^{} n_{\rm{gas}}(x,y,z)dz \times \int_{}^{} n_{\rm{CR}}(x,y,z) dz$. In the framework of the 3D likelihood analysis and the three-dimensional distribution of the CRs and gas presented in this paper, we quantified possible deviations from a steady-source injection model, using e.g. $S(x, y) = A \int_{}^{} n_{\rm{gas}}(x,y,z) \times n_{\rm{CR}}(x,y,z) dz$. We parametrized the CR gradient $\propto 1/r^{\alpha}$ and computed the likelihood profile as a function of $\alpha$, leaving the spectral parameters of all the components free. Figure~\ref{fig:morpho_CMZ} shows the likelihood profile with respect to $\alpha = 1$ and indicates a best-fit value of $\alpha = 1.10 \pm 0.05$. Since the CR density profile depends strongly on the assumed density and distribution of the gas, the value of $\alpha$ is particularly sensitive to the central clouds where the profile peaks. To reduce this source of uncertainty, we masked a circular region of $0.274^{\circ}$ radius centered on $(l, b)$ = (359.924\dg,  $-0.017$\dg) and recomputed the $\Delta$TS profile as a function of $\alpha$ and using the two H.E.S.S. analysis and reconstruction chains. For both analyses, the best-fit value is found to be $\alpha = 1.10_{-0.11}^{+0.10}$. Masking a smaller region (0.2\dg) shifts the best-fit value of $\alpha$ by $\pm$ 0.1 depending on the H.E.S.S. analysis chain used. One can therefore estimate systematic uncertainties in our data, which are not related to our model concerning the line-of-sight position of the gas. This yields $\alpha = 1.10 \pm 0.05_{\rm{stat}} \pm 0.10_{\rm{syst}}$, which supports the continuous injection scenario near the GC and validates the assumption of $\alpha = 1$ made in this work.

\begin{figure*}
\sidecaption
  \includegraphics[width=12cm]{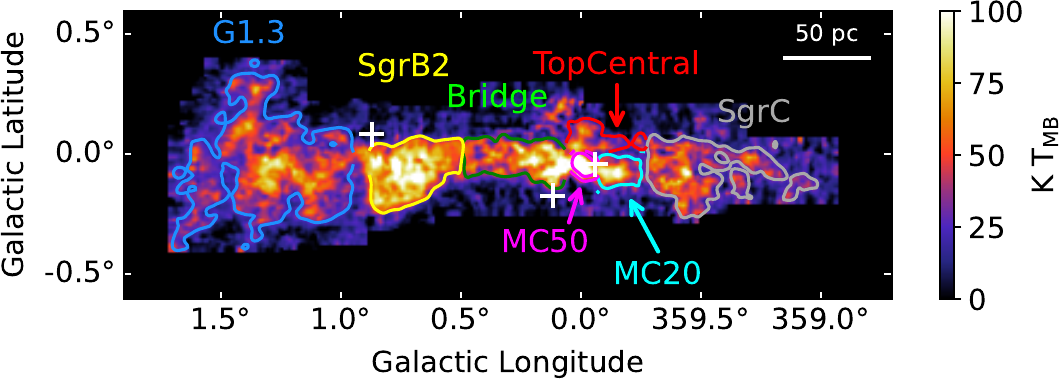}
    \caption{Velocity-integrated CS map (in brightness temperature) obtained from \cite{Tsuboi:1999}. The considered clouds are represented by different colors while the point sources are represented by the white crosses.}
    \label{fig:CSmap_Clouds}
\end{figure*}

To estimate the CR density distribution independently of the three-dimensional gas description, we split the 2D$(l, b)$ gas tracer templates into seven main molecular complexes (as shown in Figure~\ref{fig:CSmap_Clouds}) and we fitted their emission with a PL spectral model. We fitted each cloud separately leaving the spectral parameters of the Galactic large-scale and hadronic components free, as well as those of the nearby sources. Using the $\gamma$-ray luminosity of each cloud, we computed the CR density using:
\begin{equation} \label{eq:CR_lum_mass}
    \begin{split}
    w_{\rm{CR}}(> 10 \hspace{0.1cm} \rm{TeV}) & \sim \\ 1.8 \times 10^{-2}
    & \times \bigg( \frac{L_{\gamma}}{10^{34} \rm{erg s}^{-1}} \bigg) \times \bigg( \frac{10^{6}M_\odot}{M} \bigg) \hspace{0.2cm} \rm{eV} \hspace{0.1cm} {cm}^{-3},
    \end{split}
\end{equation}
with $M$ and $L_\gamma$ the mass and luminosity ($E_{\gamma} > 1$~TeV) of each cloud \citep{HESS_GC_PeVatron:2016}. 
We performed the fit and the mass estimation with different gas tracers (CS, C$^{18}$O, $^{12}$CO, $^{13}$CO and HCN). The best-fit parameters (and associated mass) of each cloud are given in Appendix~\ref{appendix:clouds_param}. We note that a curved spectral model is not significantly preferred over the PL model for any cloud, likely due to a lack of statistics.

\begin{figure*}[t!]
    \centering
    \begin{minipage}{.47\textwidth}
        \centering
        \includegraphics[scale=0.48]{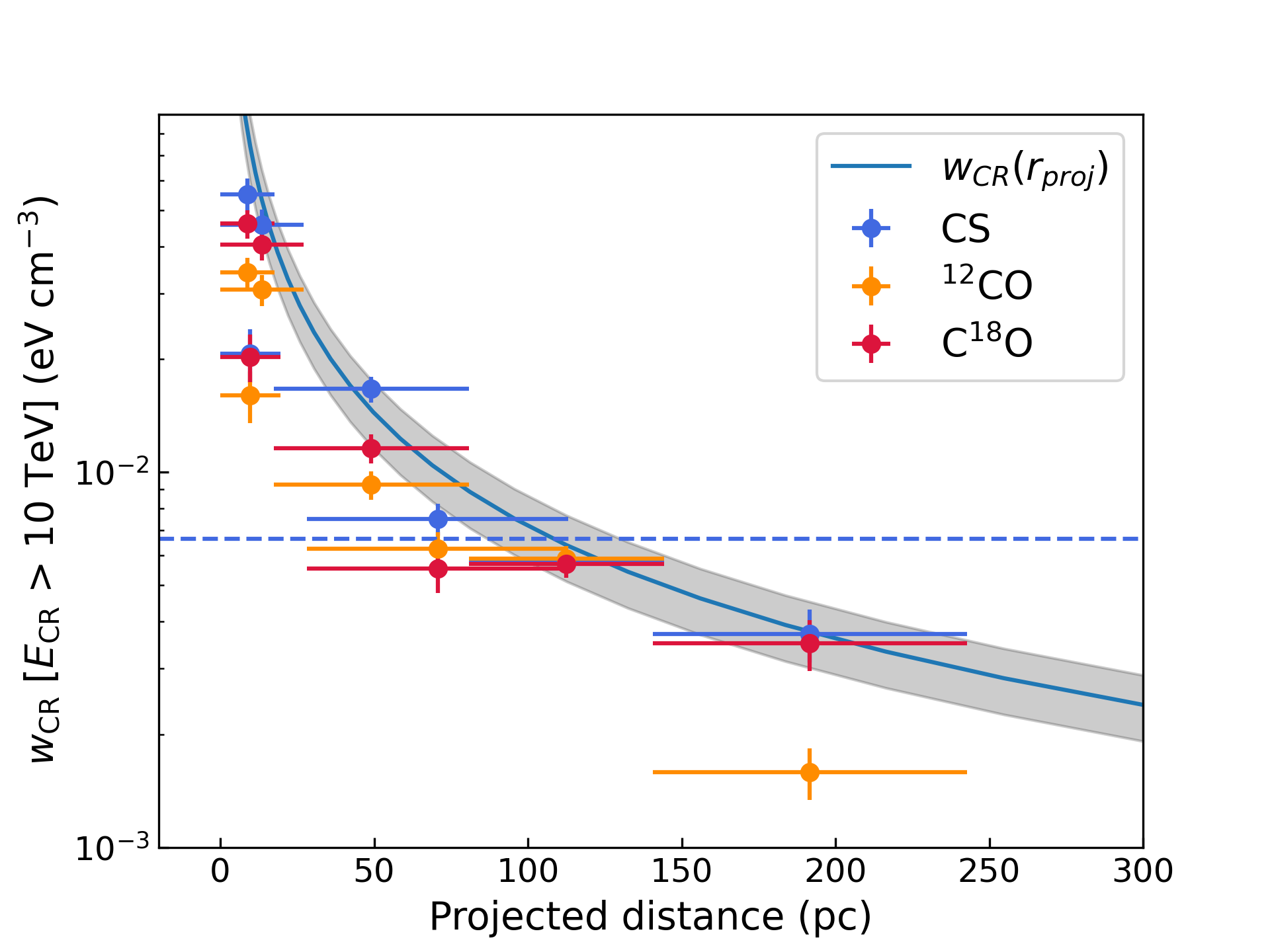}
    \end{minipage}
    \begin{minipage}{0.47\textwidth}
        \centering
        \includegraphics[scale=0.38]{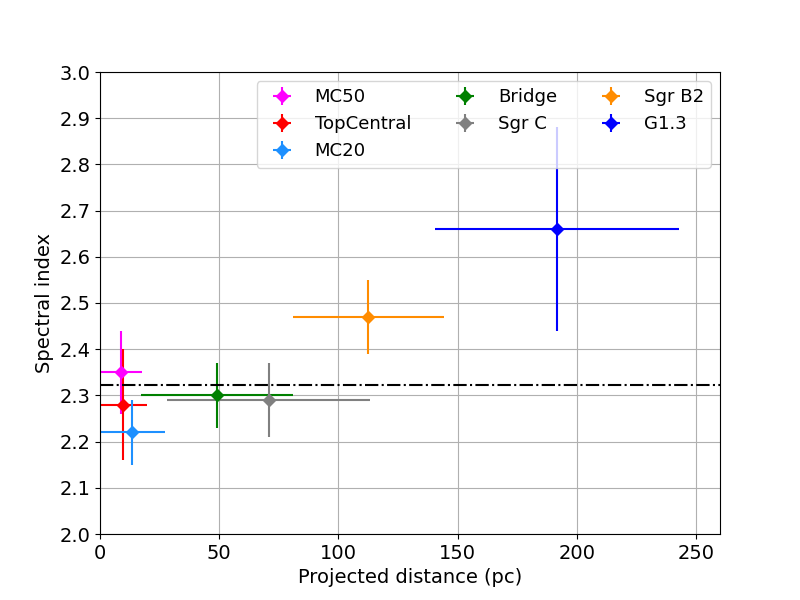}
    \end{minipage}
    \caption{(Left) CR densities derived from spectral fitting of individual 2D$(l, b)$ clouds using CS, $^{12}$CO and C$^{18}$O gas tracers as a function of the projected distance to the GC ($r_{\rm{proj}}$). The fit for a constant CR density is shown by the dashed blue line for the CS gas. The blue curve represents the CR density radial profile $w_{CR}$ deduced in Section~\ref{sec:GC_ridge}. The shaded region shows the uncertainty for a range of CMZ total mass of $(2-3) \times 10^7$\,M$_\odot$. (Right) Best-fit spectral index obtained for each cloud with respect to the projected distance from the GC. The fit for a constant spectral index is shown by the dashed black line.}
    \label{fig:CR_density_profile_and_Spec}
\end{figure*}

Figure~\ref{fig:CR_density_profile_and_Spec} (left) shows the CR density as a function of projected distance to the GC for the different clouds. For visibility, we only show the results for C$^{18}$O, $^{12}$CO and the CS gases (the other tracers led to similar profiles). For all gas tracers, the CR density profiles show an excess near the GC, and a constant density profile is ruled out at a 15$-$18$\sigma$ (CS $-$ $^{12}$CO) confidence level and 9$-$12$\sigma$ when considering only clouds with projected distance $r_{\rm{proj}} > 30$\,pc.  Figure~\ref{fig:CR_density_profile_and_Spec} (left) also shows the CR density profile $w_{CR}$ derived in Section~\ref{sec:GC_ridge}) as a function of radial projected distance. Comparing the CR profile points with $w_{CR}(r)$ requires assuming that the average distance of the molecular cloud complexes to the GC is comparable to their projected distance on the plane of the sky (the gas model we built suggests this approximation is acceptable for most of the clouds considered). We therefore obtain similar CR density profiles when using either a three-dimensional description of the CS gas or a 2D spatial template with different gas tracers.

    \subsection{Spectrum across the CMZ}

If CRs are injected continuously by a steady source at the GC and diffuse homogeneously and isotropically, no spectral variations are expected across the CMZ. In the impulsive injection scenario of protons, since the highest energy particles escape and diffuse faster than the lower ones, we expect to see them farther away from the GC. That would lead to a harder spectrum in the outer parts compared to the inner ones, and to an energy-dependent morphology (less peaked at higher energy).

From the best-fit spectrum derived for each cloud (given in Appendix~\ref{appendix:clouds_param}), we plot in Figure~\ref{fig:CR_density_profile_and_Spec} (right) the variations of the spectral index with respect to the projected distance to the GC. Although a slight hardening towards the GC is suggested, a fit of a constant spectral index gives $\Gamma = 2.322 \pm 0.002$ and data points deviate from this constant value at less than 1$\sigma$ confidence level ($p$-value of 0.19). Fitting the data points with a linear function yields a best-fit of $(2.25 \pm 0.05) + ( (1.58 \pm 0.77) \times 10^{-3})r_{\rm{proj}}$ with a $p$-value of 0.61, which is not  statistically preferred over the constant model. Thus, no significant spectral variation is observed across the CMZ, consistent with a continuous injection scenario.

\section{Discussion}\label{sec:discussion}

In this section, we discuss the possible origin of the CR injection including location and energy budget and assess a potential contribution of the Arches and Quintuplet clusters. We also examine propagation scenarios capable of producing the observed spectral curvature of the GC ridge emission and discuss the contribution to \hessGCsource of CR interaction with the CNR. Finally, we discuss constraints on the line-of-sight positions of CMZ clouds.

    \subsection{On the origin of the particle injection}

Despite the large number of potential accelerators in the region, the GC ridge is under luminous\footnote{Taking a supernova (SN) every 10 kyr and 10\% of each SN kinetic energy transmitted into protons, we expect a $\gamma$-ray luminosity two orders of magnitude higher than what is currently observed.}. This implies either inefficient acceleration near the GC or rapid CR escape via advection or diffusion. Taking a typical diffusion coefficient at 10\,TeV of $D_0 = (5-9) \times 10^{29}$\,cm$^{2}$\,s$^{-1}$, CRs need a time $t = \frac{R^2}{6D(E)} \sim 3-5$\,kyr to fill the inner 150\,pc, a timescale shorter or comparable to that of the potential sources discussed below.

        \subsubsection{Injection location and energy budget}

To assess the location of the injection site, we generated templates by shifting the CR injection site along Galactic longitude with 0.05\dg steps (equivalent to 7\,pc at the GC distance) between $l = [-0.2$\dg, $+0.2^{\circ}$] using the latitude and line-of-sight distance of the GC. For each fit, the spectral parameters of each component were left free and likelihood variations were computed with respect to an injection site at the position of \sgra. We found that the fit quality degrades significantly when the injection site is shifted towards positive longitudes, in particular at the position of the Arches and Quintuplet clusters ($\sim +14$\,pc) we obtained $\Delta$TS = $-150$. The best-fit longitudinal location was obtained for $l = -14.0_{-9.7}^{+4.2}$\,pc using a confidence range of 3$\sigma$ (\sgra being located at $\sim -7$ pc). Using cross-checks with alternative analysis chains as well as varying the fit region, we estimated our  systematic uncertainties at a level of $\pm$\,7 pc. Therefore, the point source injection should lie between $+1.2$\,pc and $-19$\,pc around \sgra (errors were added quadratically).

In steady state, the CR injection power of the central accelerator $\dot{Q}$ can be expressed as $w_{CR}(E,r) = \frac{\dot{Q}(E)}{4\pi D(E) r}$. Using the CR energy density derived in Section \ref{sec:GC_ridge} as $w_{CR}(>10\ \mathrm{TeV}) \sim 0.24 \times \left( \frac{r}{3\ \rm pc} \right)^{-1}\ \rm eV\,cm^{-3}$, we obtained:
$$\dot{Q}(E>10\rm\ TeV) \sim 2.3\times 10^{37} \left(\frac{D(E)}{5 \times 10^{29}\rm\ cm^2\,s^{-1}}  \right) \mathrm{\ erg\,s^{-1}}$$ 
We note that this injection power only accounts for particles with $E > 10$ TeV, and tighter constraints could be obtained with the spectrum of lower energy particles.

The hot accretion flow on the supermassive black hole \sgra is known to be radiatively inefficient and must dissipate a significant fraction of its accretion power into outflows and possibly in the form of energetic particles. Models based on the recent
Event Horizon Telescope  campaigns suggest an outflow power of $L_{\rm{outflow}} = (1.3 - 4.8) \times 10^{38}$ \,erg\,s$^{-1}$ \citep{EHT2022b}. Although $D(E)$ and the conversion efficiency from mechanical to CR power are not known, the outflow of \sgra could significantly contribute to the CR population in the CMZ.

Massive stellar clusters are also believed to efficiently accelerate particles, in particular through the SNRs they host or collective effects from stellar winds \citep{Cesarsky83:1983SSRv...36..173C, Aharonian:2019, Bykov2018:2020SSRv..216...42B, Morlino21:2021MNRAS.504.6096M, Vieu:2023}. Several measurements of $\gamma$-ray emission from regions surrounding several luminous compact star clusters have been reported in recent years \citep{Abramowski_2012, Westerlund2:Yang_2018,Westerlund1:Aharonian_2022, LHAASO:2024_Cygnus} corroborating this hypothesis. In the vicinity of the GC lie three bright and young stellar clusters: the Quintuplet and the Arches clusters ($\sim$ $3-5$\,Myr and $2-3$\,Myr) located within a projected distance of $\sim$ 25\,pc from \sgra and the Young Nuclear Stellar Cluster \citep[YNSC, $\sim$ 6\,Myr, ][]{Martins:2007, Feldmeier-Krause:2015} located in the inner pc and surrounding \sgra (with a total mass $\sim 1 \times 10^{6}$\,M$_{\odot}$ and a spatial extension evaluated to be $\sim $ 7 $\pm$ 2\,pc). Given our constraint on the injection site location, the YNSC appears as the most promising candidate among the three clusters. With a mass-loss rate of its Wolf-Rayet population evaluated to be $\sim$ 10$^{-3}$ M$_{\odot}$\,yr$^{-1}$ and a typical stellar wind velocity of $v_w \sim$ 1000\,km\,s$^{-1}$ \citep{Scherer2023}, the kinetic luminosity of the YNSC reaches $L_{w} = \frac{1}{2}$ $\dot{M}$ $v_{w}^2$ $\sim$ 3$-$4 $ \times 10^{38}$ erg\,s$^{-1}$, which would be sufficient to power the continuous CR injection.

Alternatively, if the CR injection originates from a single impulsive source like an SNR, rather than a stationary source, the resulting CR density profile would deviate from the canonical 1/r form and instead exhibit a 3D Gaussian shape. We investigated this scenario by approximating the gamma-ray ridge emission by the product of the 2D gas template by a 2D Gaussian profile (see Appendix~\ref{appendix:impulsive_scenario}). In this case, an additional extended source is required to explain the pronounced CR density excess observed within the inner 30\,pc. A candidate for an impulsive event is the SNR Sgr A East (with age $\lesssim$ 10 kyr), which could have injected sufficient energy into the CMZ to produce the observed large-scale TeV emission \citep{HESS_GC_ridge_discovery:2006}. In this scenario, the additional central contribution could plausibly arise from one or several unresolved sources in the immediate vicinity of the GC.

Another possibility for the origin of the $\gamma$-ray emission could be the population of SNRs embedded in clusters and in the disk. \cite{Jouvin_SNR_CRs:2017} shown that using a reasonable recurrence rate of SNRs, a quasi-stationary emission peaking towards the central parts of the CMZ could be reproduced. If the recurrence time of SNRs is less than the escape time (being energy-dependent), these impulsive injections could mimic a continuous source in the inner tens of pc the CMZ. In this case, the higher-energy particles would provide a flatter profile than the lower-energy ones resulting in an energy-dependent morphology. This could be probed with the next generation of Cherenkov telescopes CTAO \citep{Jouvin_SNR_CRs:2020}.

        \subsubsection{Testing an additional contribution from the Arches and Quintuplet clusters}

We want to estimate a potential subdominant contribution from the Arches and Quintuplet clusters (with a size of $\sim$ 2\,pc and $\leq$ 1\,pc and a mass of $\sim 1 \times 10^{4}M_{\odot}$ and $\sim 2 \times 10^{4} M_{\odot}$ respectively). With mass-loss rates of $\dot{M} \sim$ 8.1 $\times 10^{-4}$\,M$_{\odot}$\,yr$^{-1}$ and 2.0 $\times 10^{-4}$\,M$_{\odot}$ yr$^{-1}$, and stellar wind velocities of 670\,km\,s$^{-1}$  and 1850\,km\,s$^{-1}$ \citep{Scherer2023}, the Arches and Quintuplet have kinetic luminosities evaluated to  $1 \times 10^{38}$\,erg\,s$^{-1}$  and  $2 \times 10^{38}$\,erg\,s$^{-1}$ respectively, which are not negligible compared to the YNSC.  

We fitted the data with a model involving two injection sites at $+0.11^\circ$ and $-0.05^\circ$ in longitudes respectively for the Arches cluster and \sgra, leaving their spectral parameters free during the fit. The results of the fit implied no contribution at the position of the Arches cluster (best-fit amplitude close to zero) compared to the central region which strongly dominates the emission. We therefore derived an upper limit on its contribution, tying the spectral shape of its emission with that of the central region, and we found it to be 24\% of the CR luminosity of the central region (at a 2$\sigma$-confidence level). Since a potential contribution from the Quintuplet cluster, located further from the GC (at $+0.16^\circ$ in longitude), is expected to be even smaller than that of the Arches cluster, we did not test an additional contribution at this position. If the YNSC is the main contributor to the central emission, and if we believe that emission from clusters should follow their kinetic luminosities, this result is difficult to reconcile with the paradigm in which all massive stellar clusters contribute with similar efficiencies to the CR population observed in the CMZ, unless the Quintuplet and Arches clusters are not located in the same regions as the target material.
This would imply that these clusters are situated significantly farther along the line-of-sight from the GC ($\gtrsim 100-150$\,pc), whereas positional, kinematical, and dynamical evidence shows that they are integral parts of the CMZ structures \citep{Kruijssen2014}. Indeed their formation is naturally explained by the orbital dynamics and dense gas flows of the CMZ, including tidal compression and collisions near pericentre passages \citep{Sormani2020}, and cannot be reconciled with a scenario placing them far from the central molecular reservoir \citep{ Longmore2012, Henshaw2023, Hosek_2022}.

    \subsection{Possible mechanisms for causing the spectral curvature}

One of the main results of our analysis is the significant curvature of the GC diffuse emission spectrum. The fits to the parent CR spectrum do not significantly distinguish between the actual spectral shape assumed (LogP, ECPL, BPL). Depending on the parent CR spectral model assumed, our spectral models fits yield different proton energy scales of the spectral transition, all in the range of 50 to 150 TeV. The observed spectral softening can be attributed to different processes: it can be intrinsic to the particle acceleration site in the vicinity of the GC or it can be caused by propagation effects in the inner hundreds of pc. 

Propagation can cause such a spectral softening at the transition between two competing mechanisms driving particle transport. A possible example consists in advective and diffusive escape of accelerated particles from the CMZ. Being energy independent, advection dominates at low energy while keeping the spectral index of the CRs unchanged. The diffusion coefficient increases with energy, so that diffusion dominates above a certain energy threshold. The diffused particle spectrum is also steeper than the injected one. Diffusion-advection propagation therefore naturally predicts a break in the spectrum. In typical Galactic CR propagation, this transition and the associated spectral break occurs only at low energies ($\lesssim$ GeV). Yet, conditions in the CMZ are extreme. There are numerous indications of a strong advective wind taking place with $v_{\rm{wind}} \sim$ few hundreds of km\,s$^{-1}$ \citep{Heywood:2019, Ponti:2019, Yusef-zadeh:2023}. Such a scenario has been studied by \cite{Scherer:2024}. Advection could dominate up to the multi-TeV regime and cause the observed break only if the wind velocity is in the order of 1000\,km\,s$^{-1}$ and if the diffusion coefficient is strongly suppressed (factor of $\sim$ 10$^{-2}$) compared to typical interstellar values. However, we note that the expected profile in an advection-dominated scenario is likely different from the $1/r$ distribution we observe. If advection was isotropic, we expect a $1/r^2$ profile and a more complex one if there is a position-dependent competition between advection and diffusion. Given the extreme values and the different CR density profile expected for advection to dominate over diffusion up to multi-TeV energies, we consider such a scenario unlikely.

Another plausible mechanism is the transition from diffusive to ballistic transport \citep[e.g][]{Prosekin:2015} where the more energetic particles will escape because their mean free path is large compared to the size of the CMZ. 

To constrain the possible physical effect at play (propagation or acceleration causing the curvature), it will be necessary to study the possible evolution of the CR profile with energy to check whether the spectral transition is associated with a change in morphology. With their improved sensitivity in the multi-TeV domain, CTAO, SWGO and LHAASO/LACT may be able to test this.

    \subsection{Contribution of the CNR to the flux of \hessGCsource}\label{sec:assoc_GC_ridge_J1745}

Since the CR density profile peaks at the GC, the interaction of these particles with gas surrounding \sgra should contribute to the $\gamma$-ray flux of \hessGCsource. The level of this contribution depends on the total mass of the ambient gas as well as its spatial distribution. It also depends on the particle distribution at small scales around the acceleration site. In particular, the distribution of particle directions can be strongly affected by ballistic particle motion \citep{Prosekin:2015, HESS_GC_PeVatron:2016}.
For simplicity, we assume here that the CR energy density saturates at $\sim$ 3\,pc from the GC at a density of $w_{0}(>10\ \mathrm{TeV})\sim 0.24$\,eV\,cm$^{-3}$.

The massive structure closest to \sgra is the CNR. It is 1.5\,pc wide, its volume is estimated to be $V_{CND} = 18$\,pc$^3$. It is tilted towards the line of sight \citep{Ferriere:2012}, and its mass is poorly constrained with estimates ranging from $10^6$\,M$_\odot$ \citep{Christopher:2005} down to $10^4$\,M$_\odot$ \citep{Requena-Torres:2012, Hsieh:2021}. 

At the GC distance, assuming isotropic distribution, the expected luminosity from CR interacting with the CNR above 1\,TeV is:

\footnotesize
\begin{equation}
\begin{aligned}
    \footnotesize
    L_{\gamma,CNR}(>1\,\rm{TeV}) = &4.5 \times 10^{34}\rm\ erg\,s^{-1}\\
    &\times\left(\frac{w_{CR}}{0.24\rm\  eV\,cm^{-3}} \right) \times \left(\frac{M}{3.7\times 10^5\,M_\odot} \right) 
\end{aligned}
\end{equation}
\normalsize

This luminosity reaches the level of \hessGCsource ($L \sim 4.5 \times 10^{34}$\,erg\,s$^{-1}$) for a CNR mass of $\sim 3.7 \times 10^5$\,M$_\odot$ that would be compatible with the mass range estimates. If the CNR mass has the lower value of  $10^4$\,M$_\odot$, the predicted luminosity of the CNR would be $\sim 3\%$ of that of \hessGCsource, which would represent a minor contribution to its flux.

    \subsection{On the line-of-sight distance of CMZ molecular clouds}\label{sec:los_positions}

\begin{figure*}[t!]
    \centering
    \includegraphics[scale=0.5]{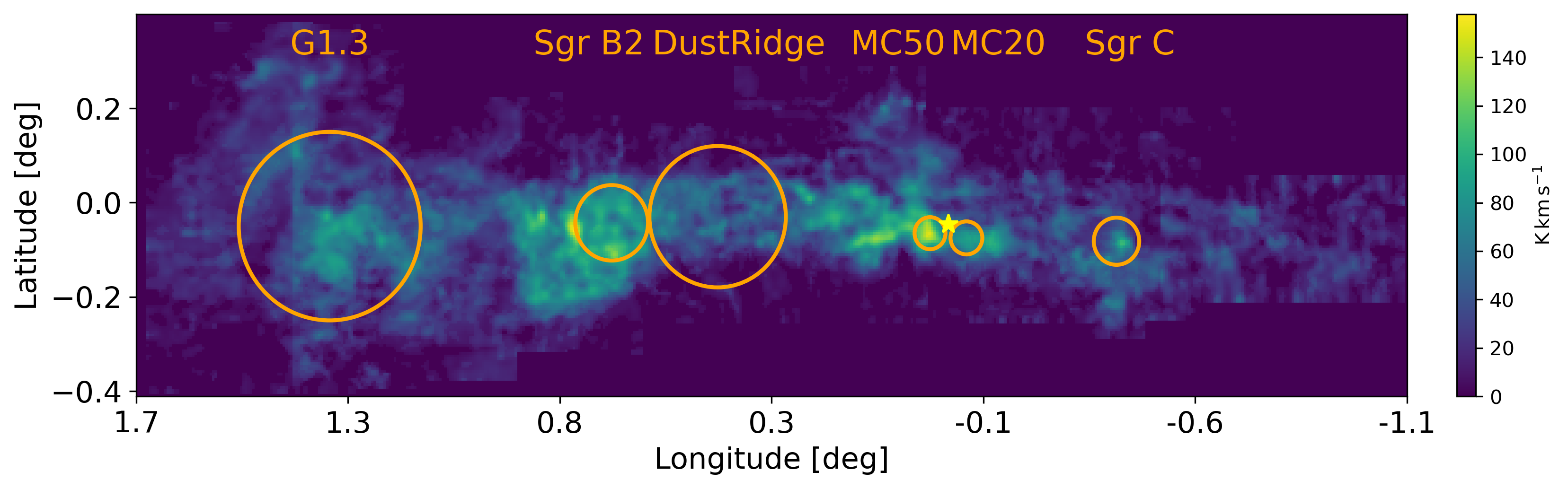}
    \includegraphics[scale=0.45]{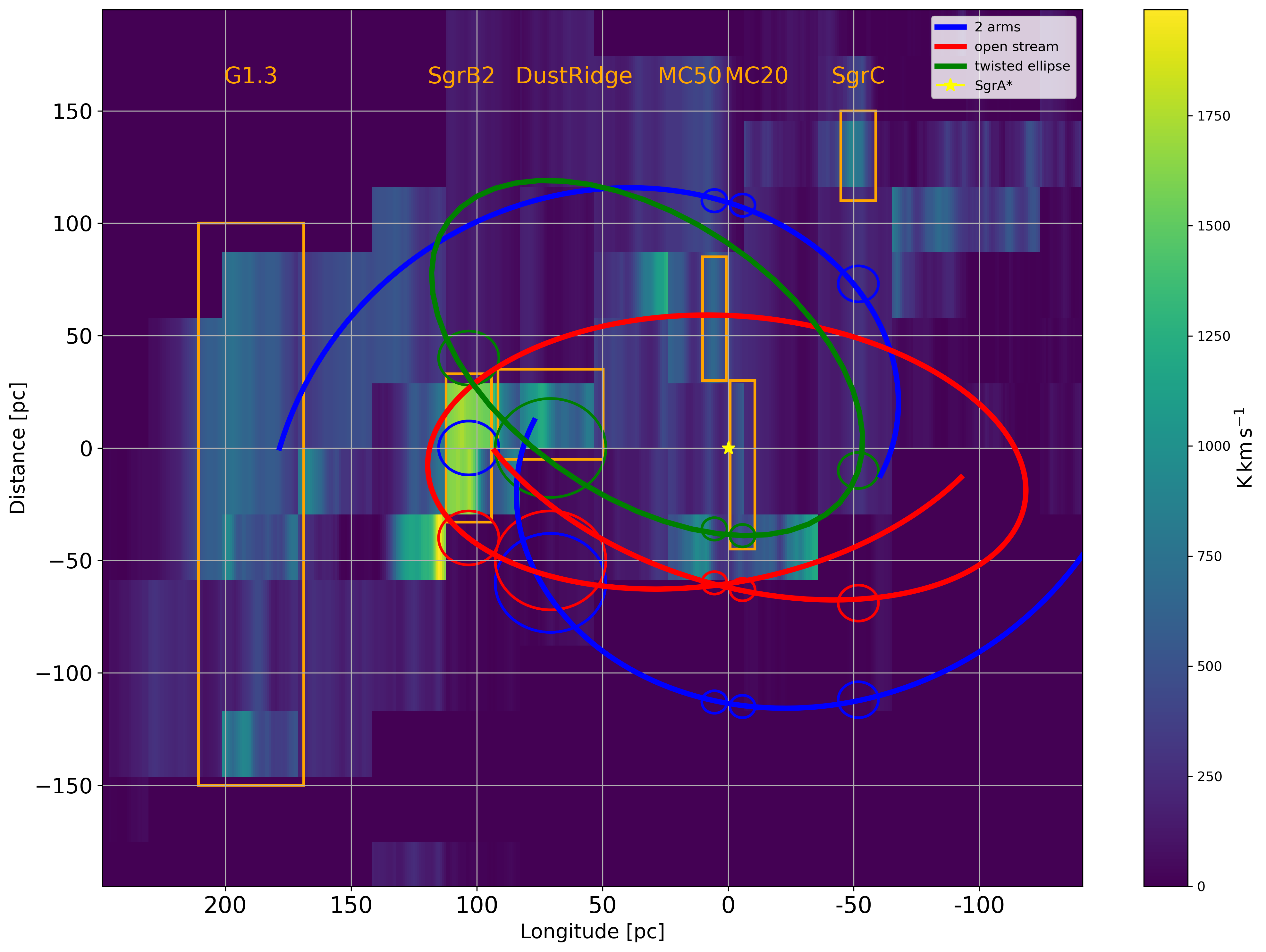} 
    \caption{$n_{gas}(x,y,z)$ cube (in units of K\,km\,s$^{-1}$) is integrated along the line of sight to produce the $(l,b)$ map (upper panel), and integrated along the latitude axis to obtain a face-on view in the $(l,z)$ plane (lower panel), where distances are expressed in pc. Both panels cover a similar longitude range. In the lower panel, each grid square represents an area of 50 $\times$ 50\,pc.
    The primary structures (the densest) of CMZ are clearly visible in both figures and are highlighted by orange circles or rectangular boxes. The predicted positions of Sgr~B2, Dust Ridge, MC50, MC20 and Sgr~C by the three other models (\textit{Two Arms}, \textit{Twisted Ellipse} and \textit{Open Streams}) are reported on the face-on view. The yellow star in both panels represents the position of \sgra.}
    \label{fig:CSmapSawada}
\end{figure*}

In this section we discuss the line-of-sight position of the major CMZ molecular clouds in the light of our new results. In Figure~\ref{fig:CSmapSawada}, the bar-like elliptical structure of the face-on map of the $n_{gas}(x,y,z)$ cube used in this work (see Section~\ref{sec:template_GC_ridge} for a description) is clearly visible, with the major axis inclined by 70$^\circ$ relative to the observed line of sight such that the Galactic eastern end appears closer to us \citep{Sawada:2004}. We note that the individual clouds structures in the $n_{gas}(x,y,z)$ cube are largely extended along the line of sight. For an extended and diffuse structure such as G1.3, this extension seems plausible, but for more compact structures such as Sgr B2 or MC 20\,km\,s$^{-1}$ / MC 50\,km\,s$^{-1}$ clouds (MC20/50 thereafter), the 30\,pc resolution of Sawada's model appears limiting and leads to unrealistic dispersion along the line of sight. It is noteworthy that despite the significant uncertainty regarding the molecular cloud mass distributions along the line of sight in the $n_{gas}(x,y,z)$ cube, it successfully reproduces the H.E.S.S. data with good accuracy in the context of a stationary model of CR injection at the GC.

In the following we compare the expected $\gamma$-ray emission of the clouds taking line-of-sight distributions from the three main dynamical models currently discussed in the literature \citep[and summarised in ][]{CMZReview2016:2016MNRAS.457.2675H, Henshaw2023} with those from the model of \cite{Sawada:2004}. We then exclude some positions based on the limits of compatibility with the H.E.S.S. $\gamma$-ray data, using the $n_{gas}(x,y,z)$ cube as a reference. 

Figure~\ref{fig:CSmapSawada} reports the positions of molecular structures for each model. 
One of them characterizes the CMZ with a pattern of two point-symmetric spiral arms around \sgra, that combine to create a ring-like structure with a radius of approximately 120 pc \citep[\textit{Two Arms} hereafter,][]{sofue1995galacticcentermoleculararms, TwoArms_Sofue2022:10.1093/mnras/stac2243}. One arm associates G1.3 with possibly MC20/50 and Sgr~C while the second arm could contain Sgr~B2, the Dust Ridge (a structure thought to be physically associated with Sgr~B2), MC20/50 and Sgr~C.
The second model interprets the CMZ molecular clouds as being continuously distributed on a closed ellipse (\textit{Twisted Ellipse} hereafter) with a vertical oscillation reproducing the infinite-shaped distribution identified in Herschel observations \citep{Molinari:2011ApJ...735L..33M}. In this model the Dust Ridge, Sgr C and MC20/50 clouds are located in the closest part of the ellipse, between $\sim -40$ pc and the GC distance, while Sgr B2 is positioned $\sim 50$\,pc beyond the GC. 
Finally, an alternative explanation proposed by \cite{OpenStream_CMZ:2015MNRAS.447.1059K}, suggests that the gas streams are disconnected and their orbits are open rather than closed (\textit{Open Streams} hereafter). This orbital solution places the five clouds in front of the GC, at a Galactocentric radius of approximately 60\,pc. As shown in Figure~\ref{fig:CSmapSawada}, none of these three dynamical models fully align with the cloud distribution in the $n_{gas}(x,y,z)$ cube (built without any kinematic assumptions), highlighting the diversity of possibilities.

MC20 and MC50: The positions of MC20 and MC50 have been the subject of high debate for many years, primarily due to projection effects caused by their close proximity to the line of sight of \sgra, the CND, and Sgr A East. Additionally, dynamical models systematically place them at large line-of-sight distances, typically between 40 and 120\,pc from the GC, while other observational data place them less than 10\,pc from \sgra \citep{Henshaw2023}. Indeed MC50 is known to interact with the SNR Sgr A East \citep{MC20-MC50Genzel90:1990ApJ...356..160G} within a radius of 4$-$5 pc from the GC. MC50 is connected to MC20 via a molecular belt suspected to be in interaction with the CND \citep{MC50-20_CND:Hsieh_2017}. Most studies conclude that these two clouds are located on either side of the CND, with evidence indicating that MC20 lies in the foreground of \sgra, as it appears as a dark feature against the 2 $\mu$m emission from the YNSC \citep{Guesten_Henkel1983:A&A...125..136G}. In this work, the $n_{gas}(x,y,z)$ cube spreads MC20 and MC50 over large distance ranges. We estimated that the averaged distribution of the $n_{gas}(x,y,z)$ cube is roughly equivalent to positioning two compact 5 parsec radius clouds at a distance from the GC between 20\,pc and 35\,pc for MC20, and between 40\,pc and 55\,pc for MC50 (the latter being situated further in the model), in term of compatibility of $\gamma$-ray flux prediction with our data, allowing a maximal variation of 30\% in  the flux measurement. Outside these intervals, the calculated $\gamma$-ray flux gradually deviates from our data, and this drift is particularly sensitive to distance variations due to the quasi-alignment of the structures with the GC line of sight. Consequently, MC20/50 are unlikely located at such a large distance than that predicted by the \textit{Two Arms} model ($d \gtrsim 100$\, pc). This is consistent with the recent analysis by \citet{Yan_2017}, which suggests more compact spatial extensions and locations closer to \sgra for these two clouds. 

Sgr C: This cloud has been identified as connected to the contiguous stream associated with the MC20/50 clouds by several studies but its position is highly debated \citep{Henshaw2023}. As shown in Figure~\ref{fig:CSmapSawada}, current models place it at a large diversity of distances ranging from 0\,pc to 110\,pc, in the background or in the foreground of \sgra. The $n_{gas}(x,y,z)$ cube places Sgr C at an average distance of $+130$\,pc (spread between $+110$\,pc and $+150$\,pc) in the background of the GC, which is in good agreement with the low $\gamma$-ray emissivity of this region despite a projected distance to the center of only 45\,pc. For instance, reducing  Sgr~C line-of-sight distance from 130\,pc to 90\,pc (respectively 75\,pc) would imply an increase of 36\% (57\% respectively) in the $\gamma$-ray flux. Given our sensitivity at this position, we estimated that line-of-sight distances $d \lesssim 90$\, pc are disfavored by our data for the core of Sgr~C as defined in Figure~\ref{fig:CSmapSawada}, thus putting the \textit{Twisted Ellipse} and \textit{Open Streams} models in tension.

Sgr B2: Sgr B2 is the most studied structure in the CMZ and has one of the best-constrained distance. It is consistently placed by the three dynamical models cited above within $\pm (50-60)$\,pc along the line of sight from the GC, either slightly in the foreground or slightly in the background. Although the limited resolution of the $n_{gas}(x,y,z)$ cube causes the structure to appear extended from $-30$\,pc to $+30$\,pc, it locates the bulk of its mass predominantly in the background of the GC. Given its projected distance to the GC of 90\,pc, the exploration of the possible positions of Sgr B2 along the line of sight leads to a distance variation to the GC between 90\,pc and 100\,pc. This results in a variation in term of encountered CR density and subsequently in $\gamma$-ray flux, of the order of only a few percent (2\%), a measurement which is typically beyond the sensitivity level of our data and prevents us from adding any further constraints on this cloud distance.

Dust Ridge: The dust ridge forms a part of a contiguous stream extending from a projected distance of $\sim 50$\,pc from the GC to Sgr~B2, with which it is thought to be physically associated. Although its visibility in IR extinction tends to place it in the foreground of the GC, in the $n_{gas}(x,y,z)$ cube it appears to be mostly located in the background of the GC. Current models predict a wide range of distances for this structure, varying from $-100$\,pc to 0\,pc. Limiting any gamma-ray flux variation to a maximum of 30\% of its value measured in this region, favors any line-of-sight distance lower than $\pm 60$\,pc, at least for the innermost region. 

G1.3: Unlike the other structures, the G1.3 complex extends over a large distance interval along the line-of-sight. Located on the edge of the CMZ in projection, its possible connection with it is still debated. The CR density inferred from the H.E.S.S. data is lower compared to the central regions. If we compare this value with the relatively low CR densities predicted by the stationary model along the line of sight, we conclude that the complex should not extend far beyond 250\,pc in radial distance from the GC, which clearly favors a connection with the CMZ.

To conclude, we qualitatively discussed the position of some clouds in the CMZ under the assumption of a stationary CR injection near the GC. With more data, better angular resolution and reduced systematics, CTAO (or LHAASO/LACT) might be able to constrain more accurately the densest cloud distances and  place stronger constraints on 3D CMZ models.

\section{Summary}

With increased statistics compared to previous publications and a 3D likelihood fitting approach with a self-consistent modeling of the region, we analyze the diffuse emission in the GC region with H.E.S.S. We place our study in the framework of a continuous injection scenario near the GC with accelerated particles diffusing through the CMZ. For the line-of-sight gas distribution, we rely on a three-dimensional description of the gas based on observational data and we convolve it with a CR density profile ($\propto 1/r$ for a continuous injection) to model the $\gamma$-ray emission from the CMZ (referred to as the GC ridge).

The multi-component model and the accounting for our main systematic errors permitted by the 3D likelihood analysis, allow us to derive for the first time the intrinsic spectrum of the point sources in the region, as well as the one of the GC ridge. We give the best-fit position and spectrum of HESS~J1745$-$290, G0.9+0.1 (detecting for the first time a significant cutoff at $E \sim 5$ TeV), HESS~J1746$-$285 and HESS~J1741$-$302, as well as the best-fit spatial extension of the Galactic large-scale component - a combination of the Galactic diffuse emission and unresolved sources. The best-fit spectral models of the GC ridge show a significance preference for a curvature or a cutoff ($>$ 3$\sigma$), implying a spectral transition near $E \sim 10-20$\,TeV, which is also consistent with the results from other experiments. This leads to a break in the proton spectrum at around $E_b \sim 15-50$\,TeV or a proton cutoff energy of $E_{\rm{cut}} \sim 50-150$\,TeV. We also test potential deviations from a steady-source injection scenario, fitting the CR radial index and deriving spectral indexes across the CMZ. We find that results are consistent with a continuous injection source near the GC with \bd{$\alpha = 1.10 \pm 0.05_{\rm{stat}} \pm 0.10_{\rm{syst}}$}, and no significant spectral variations within the CMZ.

We investigate the potential origin of the GC ridge emission based on energy budget arguments and we use our data to constrain the CR injection site location. Our results indicate that the injection site is confined within the central few pc of the Galaxy, and they strongly disfavor a scenario in which the main emission originates from the vicinity of the Arches and Quintuplet clusters. We also estimate the parameters needed to explain the curvature with propagation effects rather than acceleration itself, and we show that a competition between advection and diffusion can hardly explain a curvature at such high energy. Besides this, we estimate that a potential contribution of the CNR to the emission of \hessGCsource is possible given its mass range estimates and the corresponding expected $\gamma$-ray flux. Finally we derive qualitative constraints on the line-of-sight distance of the clouds by allowing a maximum of 30\% of variations in the $\gamma$-ray flux compared to the best-fit model found in this work. The tighter constraints are obtained for the MC~20\,km\,s$^{-1}$ and MC~50\,km\,s$^{-1}$ clouds, Sgr~C and the Dust Ridge, that challenge some proposed positions from different dynamical models.

\begin{acknowledgements}
The support of the Namibian authorities and of the University of Namibia in facilitating the construction and operation of H.E.S.S. is gratefully acknowledged, as is the support by the German Ministry for Education and Research (BMBF), the Max Planck Society, the Helmholtz Association, the French Ministry of Higher Education, Research and Innovation, the Centre National de la Recherche Scientifique (CNRS/IN2P3 and CNRS/INSU), the Commissariat à l’énergie atomique et aux énergies alternatives (CEA), the U.K. Science and Technology Facilities Council (STFC), the Polish Ministry of Education and Science, agreement no. 2021/WK/06, the South African Department of Science and Innovation and National Research Foundation, the University of Namibia, the National Commission on Research, Science \& Technology of Namibia (NCRST), the Austrian Federal Ministry of Education, Science and Research and the Austrian Science Fund (FWF), the Australian Research Council (ARC), the Japan Society for the Promotion of Science, the University of Amsterdam and the Science Committee of Armenia grant 21AG-1C085. We appreciate the excellent work of the technical support staff in Berlin, Zeuthen, Heidelberg, Palaiseau, Paris, Saclay, Tübingen and in Namibia in the construction and operation of the equipment. This work benefited from services provided by the H.E.S.S. Virtual Organisation, supported by the national resource providers of the EGI Federation.
\end{acknowledgements}

\bibliographystyle{aa}
\bibliography{biblio_GC_HESS}

\begin{appendix}
\newpage
\onecolumn

\section{Systematic uncertainties}

    \subsection{Spatial and spectral models used for the diffuse emissions}\label{appendix:syst_spatial_spec_models}

Performing a 3D$(l, b, E)$ likelihood fit implies (a priori) a model selection. In particular, it requires choosing a proper spatial model to describe the emission from extended sources. We investigated that using an alternative spatial model to describe the GC ridge, like a 2D Gaussian CR gradient in the CMZ \citep[as in ][]{HESS_Diffuse_emission:2018}, led to a similar best-fit spectrum (see Appendix~\ref{appendix:impulsive_scenario}). We also verified that our results are not affected by the spatial model used for the Galactic large-scale emission, using spatial $\gamma$-ray templates provided by the HERMES code \citep{HermesCode}, assuming a homogeneous CR distribution and a radially-dependent diffusion coefficient using HI and CO gas distributed into rings with different Galactocentric radii based on \cite{LAT_GIEM:2016} and \cite{Galview1} (and excluding the CMZ region). We repeated our analysis using these physical models and found compatible results for the best-fit position of the point sources and the best-fit spectra of all the components. Finally we tested several spectral models for the Galactic large-scale and GC ridge emissions and evaluated their impact on the results using data and simulations. First, we checked on data that the different spectral models used for the GC ridge do not have any significant impact on the point-source spectral measurements. Then we estimated the impact of the Galactic large-scale emission spectral shape on the GC ridge spectrum measurement. The results show that any spectral shape including a curvature to model the Galactic large-scale component affects the GC ridge curvature only marginally (similar significance and best-fit values) compared with other systematic errors. In consequence, we did not add any contribution due to spatial or spectral model uncertainties to the calculation of systematic errors. However, to assess the robustness of the curvature measured in the GC ridge spectrum, we also tested an extreme scenario in which the Galactic large-scale component's curved spectral shape is replaced with a simple power law, despite the significant curvature detected in its spectrum. In this case, the curvature in the GC ridge spectrum is even more significant ($> 5\sigma$), as shown in Table~\ref{tab:delta_ts_values}, regardless of the spatial modeling of the large-scale emission (using the uniform-longitude model or HERMES template). Moreover, the cutoff energy of the GC ridge is shifted towards lower value, as reported in Table~\ref{tab:best_fit_values}, which likely compensates for the absence of a spectral curvature in the large-scale component (clearly favored by our data). Therefore our results reporting a detection of curvature of the GC ridge emission at more than 3$\sigma$ level are conservative and based on the best modeling of the region.

\begin{table}[h!]
\caption{$\Delta$TS values obtained for the curved spectral models (LogP or ECPL) with respect to the PL model for the GC ridge spectrum, when using different spatial and spectral modeling of the large-scale emission.}
\centering
\begin{tabular}{lcccc}
\hline
\hline
\multicolumn{1}{c}{} & Uniform-long $\times$ LogP  &  HERMES $\times$ LogP & Uniform-long $\times$ PL  & HERMES $\times$ PL \\
\hline
$\Delta$TS$_{\rm{PL \rightarrow LogP}}$   &  14.1 & 15.8 & 25.1	 &  26.1 \\ 
$\Delta$TS$_{\rm{PL \rightarrow ECPL}}$  & 14.6  & 16.4 & 26.4 & 26.9 \\ 
    \hline
\end{tabular}
\tablefoot{The spatial models used for the large-scale emission are the uniform-longitude model (with a latitude Gaussian $\sigma$ extent) and the template computed with the HERMES code.}
\label{tab:delta_ts_values}
\end{table}

\begin{table}[h!]
\centering
\begin{tabular}{lcccc}
\hline
\hline
\multicolumn{1}{c}{} & Uniform-long $\times$ LogP  &  HERMES $\times$ LogP & Uniform-long $\times$ PL  & HERMES $\times$ PL \\
\hline
$\alpha$   &  $2.14 \pm 0.06$ & $2.10 \pm 0.06$ & $2.06 \pm 0.06$ & $2.04 \pm 0.07$ \\ 
$\beta$   &  $0.12 \pm 0.03$ & $0.12 \pm 0.03$ & $0.15 \pm 0.03$ & $0.16 \pm 0.04$ \\
\hline
$\Gamma$ &  $2.14 \pm 0.06$  &  $2.10 \pm 0.06$    &  $2.05 \pm 0.07$   &  $2.04 \pm 0.06$     \\
$E_{\rm{cut}}$ (TeV)  & $18.42 \pm 6.08$  & $17.62 \pm 4.94$ & $12.53 \pm 3.51$ & $12.56 \pm 3.09$ \\ 
\end{tabular}
\caption{Best-fit spectral parameters of the GC ridge ($\alpha$, $\beta$ and $\Gamma$, $E_{\rm{cut}}$ are for the LogP and ECPL model respectively) when using different spatial and spectral modeling of the large-scale emission.}
\label{tab:best_fit_values}
\end{table}

    \subsection{IRF and background systematic error estimations with Monte Carlo simulations}\label{appendix:syst_MC}

We performed Monte-Carlo simulations to evaluate the systematic errors on the spectral parameters fitted in Sections~\ref{sec:HESS_sources} and ~\ref{sec:GC_ridge}. Using \textsc{gammapy}, we used the reduced dataset cube in FITS format (including counts cube, PSF cube, exposure cube and background model) and the best model fitted, as a basis to perform Monte-Carlo simulations and create fake events cubes.  We introduced several systematic effects on the IRFs and background model in the simulations. For the IRFs, we considered uncertainties on the collection area estimation and energy reconstruction. Uncertainties on the size of the PSF and on the pointing position could also affect the reconstruction and have an impact on the morphology, but since we have shown that the precision of the spatial modeling has no critical impact on our spectral results, we choose to not investigate these aspects here. We included an energy bias on the order of 10\% in the source spectral models, a dispersion of the collective area normalization and slope of 10\% and an average energy shift of 10\% (see Table~\ref{SYST_parameters}). The order of magnitude of these uncertainties were inspired by the work in \cite{Nigro_2019}. For the background, we introduced a typical dispersion of the global normalization of 10\%, and added a dispersion in the spectral count slope of 0.05, of the order of magnitude of typical values measured in \cite{Mohrmann:2019}. As a second step we refitted the best-fit model on the simulated data cubes to see how each fitted parameter can be altered by those uncertainties. We iterated this procedure 500 times for each spectral model shape and for each component of the model.

\begin{figure}[h!]
    \centering
        \includegraphics[width=0.55\textwidth]{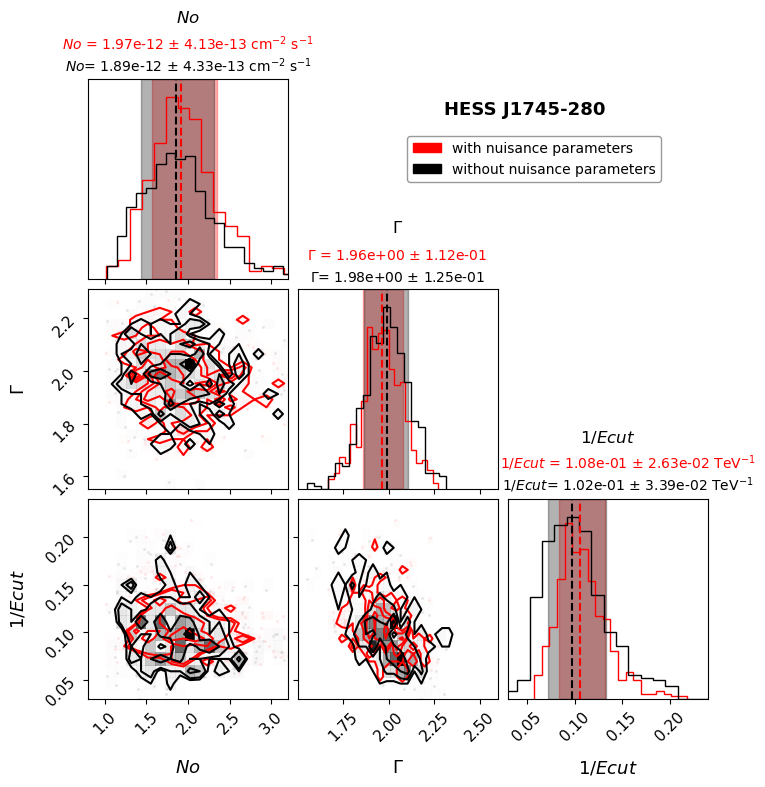}
        \includegraphics[width=0.55\textwidth]{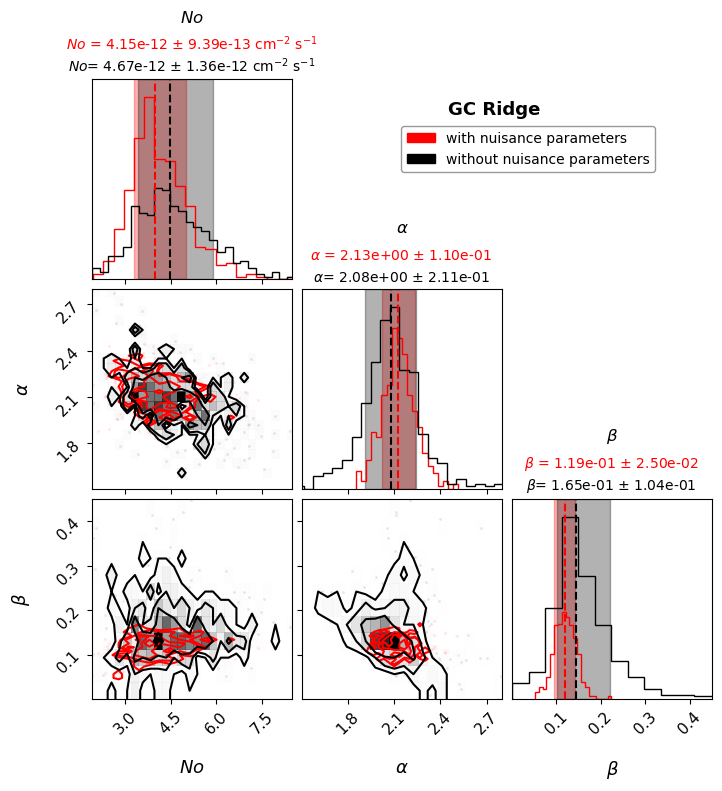}
    \caption{Corner plot showing the fitted spectral parameters of \hessGCsource and the GC ridge from 500 simulated datasets. Each simulation includes random systematic deviations in the IRFs and background models, as described in Table~\ref{SYST_parameters}. The 2D contour plots represent the correlations between parameters, while the 1D histograms display their marginal distributions. Results obtained when fitting with and without systematic nuisance parameters are shown in red and black, respectively. The mean and standard deviation of each distribution are indicated above the diagonal.}
    \label{fig:simulations}
\end{figure}

We then plotted the model parameter dispersion distribution and estimate confidence errors for each parameter. An example of the resulting distribution of fit parameters for \hessGCsource and the GC ridge emission can be seen in Figure~\ref{fig:simulations}. Note that the parameter dispersions seen on Figure~\ref{fig:simulations} correspond to the total errors including both statistical and systematic effects taken into account in the simulation.
By quadratically subtracting the contribution of statistical errors (estimated from the likelihood minimization) to the standard deviation of the distributions, we derive the systematic errors for each spectral parameter and quote it in Tables~\ref{tab:best_fit_spec} and ~\ref{tab:CMZ_spec} 
\footnote{To be precise, given that part of the uncertainties of the IRFs and background are already taken into account in the fit procedure by the introduction of nuisance parameters, the statistical errors reported in Tables~\ref{tab:best_fit_spec} and ~\ref{tab:CMZ_spec} in fact already include part of the systematic errors. What we call “systematic errors” might therefore be better termed as “residual systematic errors”.}.
\begin{table}[h!]
\caption{List of parameters used to introduce systematic uncertainties in the IRFs and the background model.}
\small\centering
\begin{tabular}{lcc}
\hline
\hline
     &     Parameters         &  Variation    \\
    \hline
    IRFs   &   $E_{\rm{bias}}$	 &    $\mu=1$ , $\sigma=0.1$ \\ 
           &   $\rm{Aeff}_{\rm{bias}}$ - norm     &    $\mu=1$, $\sigma=0.1$ \\
          &   $\rm{Aeff}_{\rm{bias}}$ - tilt     &    $\mu =0$ , $\sigma = 0.1$ \\
        \hline
       Background        &   norm   &    $\mu=1$ , $\sigma=0.1$ \\ 
                         &   tilt   &    $\mu=0$ , $\sigma=0.05$ \\
    \hline
\end{tabular}
\tablefoot{For each parameter, the properties of the corresponding sampling distribution are provided. In all cases the distribution of biased parameters follow a Gaussian with an average value $\mu$ and a standard deviation $\sigma$.}
\label{SYST_parameters}
\end{table}

\section{2D Gaussian gradient in the CMZ}\label{appendix:impulsive_scenario}

\begin{figure*}[h!]
        \centering
        \includegraphics[scale=0.2]{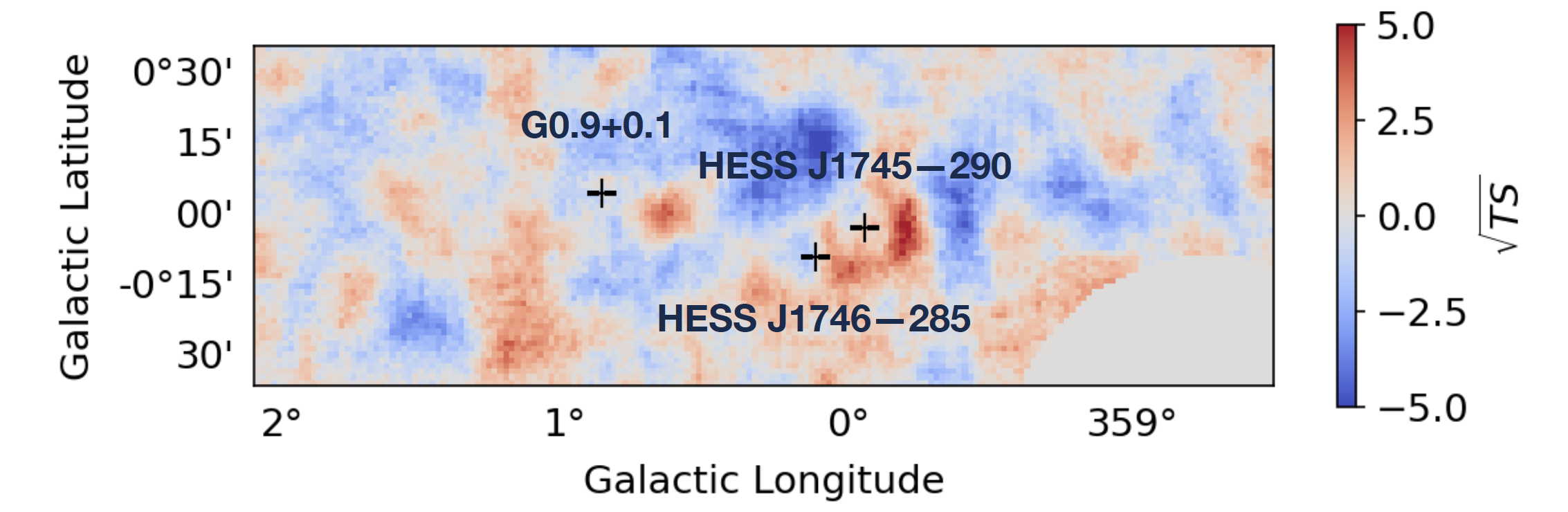}
        \includegraphics[scale=0.3]{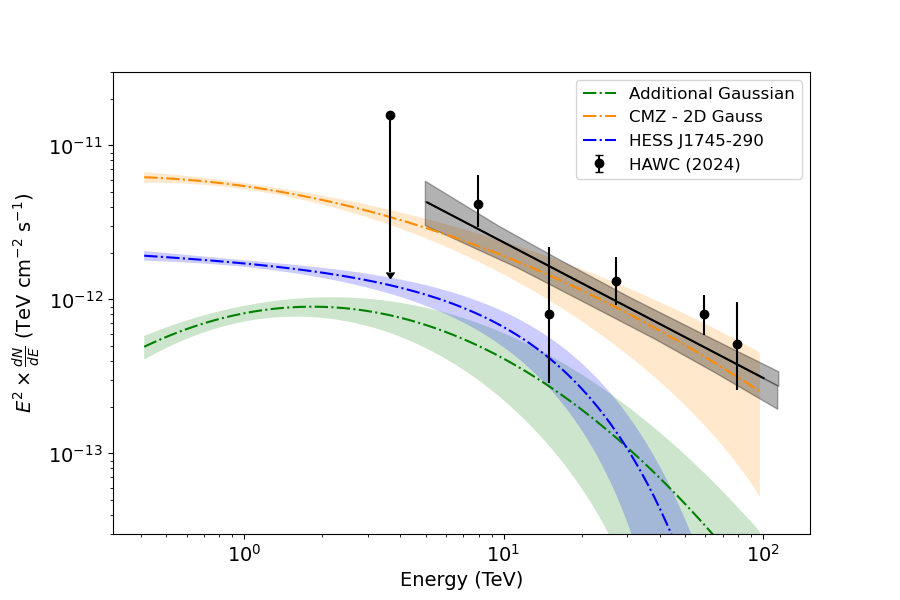}
        \caption{(Left) Residual significance map (0.4$-$100 TeV) obtained after applying a 2D Gaussian gradient within the CMZ. (Right) Best-fit spectra and SEDs of the GC ridge when applying a 2D Gaussian gradient in the CMZ and of the additional Gaussian component, together with the spectrum of \hessGCsource and the one from HAWC extracted towards the GC.}
        \label{fig:Impulsive_scenario_Sig_map_and_profile}
    \end{figure*}

In \cite{HESS_Diffuse_emission:2018}, a 2D Gaussian gradient was applied to the gas component, which can be approximately associated to a scenario in which we have impulsive injection from a single source at the GC. In this case, an additional component was needed to reproduce the peaked emission, which was placed at $l, b$ = 0\dg, 0\dg and with $\sigma$ = $0.11^{\circ} \pm 0.01^{\circ}_{\rm{stat}} \pm 0.02^{\circ}_{\rm{syst}}$. With the 3D likelihood approach and the gas distribution $n_{\rm{gas}}(x,y,z)$ used in this work, we performed the same analysis applying a 2D Gaussian gradient instead of a $1/r$ profile. We found a best-fit sigma extent of $\sigma$ = 0.73\dg $\pm$ 0.04\dg and we obtained the residual significance map shown in Figure~\ref{fig:Impulsive_scenario_Sig_map_and_profile} (left), showing the contribution from the additional central component. We modeled it by a symmetric Gaussian ($\Delta$TS = 95.7), with a best-fit sigma extent of $\sigma$ = 0.88\dg $\pm$ 0.05\dg. The spectrum of the additional component is significantly curved with $\Delta$TS (PL $\rightarrow$ LogP) = 16.0 (4$\sigma$), as the one of the GC ridge with $\Delta$TS (PL $\rightarrow$ LogP) = 9.4. The best-fit parameters obtained for the GC ridge using a logarithmic parabola model are $\alpha$ = 2.23 $\pm$ 0.06, $\beta$ = 0.10 $\pm$ 0.04 and $N_0 = (5.45 \pm 0.26) \times 10^{-12}$ \dfu, compatible with the ones obtained when applying the $1/r$ gradient. The best-fit spectra and SEDs are shown Figure~\ref{fig:Impulsive_scenario_Sig_map_and_profile} (right).

\section{Parent proton distribution}\label{app:proton_distrib}

To measure directly the parent CR particle distribution, we assumed that the differential CR density at distance $r$ and energy $E_p$ follows:

$$
\frac{dN_p}{dE_p\ dV} = n_{p,0} \left(\frac{r}{r_0}\right)^{-1} \Phi(E_p)
$$
where $n_{p,0}$ is the differential CR density at distance $r_0$ and $\Phi$ is the spectral distribution of particles. As a function of position and energy, the predicted flux of emitted $\gamma$ rays at the GC distance $D_{GC}$ is:

$$
\begin{aligned}
\frac{dN_{\gamma}}{dA\ dE_{\gamma}\ dt} 
   = &\frac{ n_{p,0} }{4\pi D_{GC}^2} \int dV\ n_{gas}(x,y,z) \times \left(\frac{r}{r_0}\right)^{-1} \times \\
   &\int dE_p\  \Phi(E_p)\  \frac{d\sigma}{dE_\gamma}(E_p, E_\gamma)\ c\\
\end{aligned}
$$
where $n_{gas}(x,y,z)$ is the gas density model and $\frac{d\sigma}{dE_\gamma}$ is the differential $\gamma$-ray production function. We used the Pythia8 parametrization from \citet{Kafexhiu:2014} as implemented in the \texttt{naima} library \citep{Zabalza:2015}. The latter assumes a constant ISM density $n_{cst}$ and a spatially-integrated differential spectrum whose amplitude $N_{p}$ is in eV$^{-1}$ and can be expressed as:

$$
\begin{aligned}
\frac{dN_{\gamma}}{dA\ dE_{\gamma}\ dt} 
   = &\frac{ n_{cst} }{4\pi D_{GC}^2} \int dE_p\ N_{p}\ \Phi(E_p)\  \frac{d\sigma}{dE_\gamma}(E_p, E_\gamma)\ c\\
\end{aligned}
$$

Using this parametrization, we performed the 3D likelihood fit to constrain the CR population spectral parameters. For the three curved models (BPL, LogP and ECPL), the best-fit spectral parameters are given in Table~\ref{tab:CMZ_proton_spec} (assuming $n_{cst}$ = 100 cm$^{-3}$ and $d = 8.2$ kpc). We can then recover the CR density spectrum normalization $n_{p,0}$ thanks to the following relation:
$$ n_{p,0} \int dV\ n_{gas}(x,y,z) \times \left(\frac{r}{r_0}\right)^{-1} = N_{p}\ n_{cst}$$
To compute the energy density, we integrated the spectral distribution of protons above 10 TeV: $w_p = \int\ dE_p E_p\ n_{p,0}\ \Phi(E_p)$. Assuming a total CMZ mass of $M_{CMZ} = (2.5 \pm 0.5) \times 10^7 \rm M_\odot$, we normalized the gas cube such that $M_{CMZ} = \int dV \ n_{gas}(x,y,z)\ m_{\rm H_2}$. Then taking $r_0 = 3$ pc, we obtained a typical CR energy density profile above 10 TeV of:
$$w_{CR}(>10 \rm\ TeV) \sim 0.24 \pm 0.05\ eV\ cm^{-3} \left(\frac{r}{3\ pc}\right)^{-1}$$

\section{Parameters of each cloud}\label{appendix:clouds_param}

We give the best-fit power-law spectral parameters obtained for each cloud in Table~\ref{tab:BF_clouds} that are used to compute the luminosity using the distances cited in Table~\ref{tab:mass_estimate} which also reports mass estimates for deriving the CR density profile as a function of the projected distance to the GC. From the literature, we considered that the mass of Sgr B2 is relatively well known and close to $\sim$ $5.17 \times 10^{6}M_\odot$. We then recovered the mass of the other clouds by using the count ratio between the respective cloud and Sgr B2, i.e. $M_{\rm{SgrC}} = \frac{\sum_{\rm{SgrC}} \rm{Cnts} \times M_{\rm{SgrB2}}}{\sum_{\rm{SgrB}} \rm{Cnts}}$. 

\begin{table}[h!]
    \caption{Best-fit spectral parameters of each individual cloud using the 2D velocity-integrated CS map. $N_0$ is evaluated at $E_0 = 1$ TeV.}
    \centering
    \begin{tabular}{l| c | c }
            \hline
            \hline
            Clouds & $N_0$ (TeV$^{-1}$\,cm$^{-2}$\,s$^{-1}$) & $\Gamma$  \\
            \hline
MC 50 km\,s$^{-1}$   &   (4.95 $\pm$ 0.50) $\times$ 10$^{-13}$        &     2.35 $\pm$ 0.09   \\
MC 20 km\,s$^{-1}$   &   (5.16 $\pm$ 0.41) $\times$ 10$^{-13}$        &     2.22 $\pm$ 0.07    \\
TopCentral          &   (2.53 $\pm$ 0.39) $\times$ 10$^{-13}$        &     2.28 $\pm$ 0.12      \\
Bridge              &   (6.46 $\pm$ 0.44) $\times$ 10$^{-13}$        &     2.30 $\pm$ 0.07 \\
Sgr C               &   (6.78 $\pm$ 0.59) $\times$ 10$^{-13}$        &     2.29 $\pm$ 0.08   \\
Sgr B2              &   (7.77 $\pm$ 0.49) $\times$ 10$^{-13}$        &     2.47 $\pm$ 0.08   \\
G1.3                &   (6.14 $\pm$ 0.87) $\times$ 10$^{-13}$        &     2.66 $\pm$ 0.22    \\
        \end{tabular}
        \label{tab:BF_clouds}
\end{table}

\begin{table}[h!]
    \caption{Mass estimate (from CS, $^{12}$CO, C$^{18}$O respectively) and projected distance of the clouds with respect to the GC using different tracers.}
    \centering
    \begin{tabular}{l|ccccc}
    	\hline
        \hline
       Clouds      &  Mass ($\times 10^{6}M_\odot$) & Projected distance (pc) & Distance (kpc) & \\
		\hline
     50 km\,s$^{-1}$ & 0.58, 0.84, 0.63 & [0, 17.6] & 8.0 & \\
     TopCentral & 0.72, 0.97, 0.87 & [0, 18.6] & 8.0 & \\
     20 km\,s$^{-1}$ & 0.73, 1.11, 0.87  & [0, 27.1] & 8.0 & \\
     Bridge & 2.17, 3.51, 2.77 & [17.4, 80.8] & 7.9 & \\
     Sgr C & 5.27, 5.00, 4.29 & [28.3, 113.1] & 8.1 & \\
     Sgr B2 & 5.17*   & [80.8, 144.2] & 7.9 & \\
     G1.3 & 5.33, 9.79, 4.69 & [140.6, 242.8]  &  7.7 & \\
     \hline
    \end{tabular}
    \vspace{0.05cm}
    \tablefoot{As the distance errors to most clouds are unknown, we used 0.05 pc for all clouds, except for G1.3 for which we used 0.1 kpc.}
    \label{tab:mass_estimate}
\end{table}

\end{appendix}

\end{document}